%% file: thdm2R.tex
\documentclass[11pt]{article}
\usepackage{graphicx}
\usepackage{latexsym}
\usepackage[fleqn]{amsmath}
\begin{document}
%\pagenumbering{roman}
%\pagenumbering{Roman}
%\pagenumbering{alph}
%\pagenumbering{Alph}
%\setcounter{page}{1}
\setcounter{equation}{0}
\setcounter{figure}{0}
\def\mib#1{\mbox{\boldmath $#1$}}
%\def\mib#1{\vec{#1}}

%%%%%%%%%%%%%%%%%%%%%%%%%%%%%%%%%%%%%%%%%%%%%%%%%%%%%%%%%%%%
%\begin{titlepage}
\begin{flushright}
July 12, 2008

OU-HET 607
\end{flushright}

\vskip1cm
\begin{center}
{\Large{\bf 
Empirically Consistent Electroweak Radiative Corrections with the Two-Higgs 
Doublet Model
}}
\end{center}

\vskip0.5cm
\begin{center}
{\large {\bf Masataka Fukugita $^{a, b}$ and Takahiro Kubota $^{b, c}$
\footnote{
e-mail: kubota@het.phys.sci.osaka-u.ac.jp
}}}
\end{center}

\vskip0.2cm
\begin{center}
$^{a}${\it Institute for Cosmic Ray Research, University of Tokyo, 
Kashiwa  277-8582, Japan}

\vskip0.2cm
$^{b}${\it
Institute for the Physics and Mathematics of the Universe, 
University of Tokyo, Kashiwa, 277-8582, Japan}

\vskip0.2cm
$^{c}${\it Graduate School of Science, Osaka University,  
Toyonaka, Osaka 560-0043, Japan}
\end{center}

\vskip1cm
\begin{abstract}
The electroweak radiative correction, which turned out to be marginal
within the standard electroweak model having the minimal Higgs sector
in view of the present experimental information, fits well the experiment
when the Higgs sector is extended to have two Higgs doublets.
We predict the range where the charged and CP odd Higgs boson masses
would lie, taking
the two CP even neutral Higgs boson masses to be degenerate
which makes the analysis in multiparameter space feasible. 
It is shown that the mass of neutral Higgs doublet
boson can arbitrarily be large consistently with the $W$ mass, 
if the
charged Higgs boson is present and it's mass lies in some appropriate ranges.
\end{abstract}
%\end{titlepage}
%

%%%%%%%%%%%%%%%%%%%%%%%%%%%%%%%%%%%%%%
\vfill\eject
%\section{Introduction}\label{Intro}
%\setcounter{equation}{0}
%\setcounter{figure}{0}
%\renewcommand{\theequation}{1.\arabic{equation}}
%\renewcommand{\thefigure}{1.\arabic{figure}}

Whereas electroweak standard model is highly successful, its structure 
of the Higgs sector is poorly explored and remains unconstrained. 
At present a uniquely useful empirical probe for the Higgs sector is 
a consistency test of the electroweak radiative corrections. The $W$ boson 
mass receives a significant electroweak radiative correction, and current 
data indicate that mass of the neutral Higgs boson be rather small in 
the standard model with the minimal Higgs sector. The current data  \cite{pdg}
\begin{eqnarray}
m_{W}=80.403 \pm 0.029\:\:\ {\rm GeV}
\label{eq:wmass}
\end{eqnarray}
require  that neutral Higgs boson mass $m_{\phi}$ be smaller than 110 
GeV, while the neutral Higgs search  at LEP results in 
\begin{eqnarray}
m_{\phi }\geq 114.4 \:\:\: {\rm GeV} 
\label{eq:smhiggsallowedregion}
\end{eqnarray}
which is already a marginal conflict with the radiative correction requirement, 
if at 1 sigma. This is particularly so, if the  mass of the top quark 
is as small as
\begin{eqnarray}
m_{t}=172.6\pm 0.8 ({\rm stat})\pm 1.1 ({\rm syst}) \:\:\: 
{\rm GeV}, 
\label{eq:topquakmass}
\end{eqnarray}
as reported recently from CDF and D0 experiments  \cite{tevatron1}, 
which is smaller than 
$m_{t}=174.2 \pm 3.3 \:\:\:{\rm GeV}, $
published in \cite{pdg}.

\begin{figure}[t]
\begin{center}
%\includegraphics{mhdr.eps}
%\hskip1cm 
%\special{epsfile=mhdr6.eps hscale=0.85, vscale=0.85}
\scalebox{0.8}[0.8]{
\input{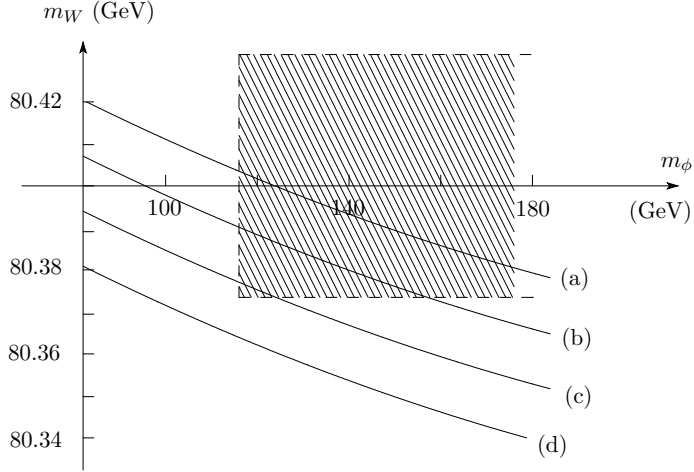}}
%\vskip7cm
\caption{W-boson mass after the electroweak radiative correction
as a function of the standard model Higgs boson mass 
$m_{\phi}$. The shade shows an empirically
allowed region, $80.432 \: {\rm GeV} > m_{W}>80.374 $ 
GeV and $m_{\phi}>114.4$ GeV. 
The solid lines correspond to 
$m_{t}=178 ({\rm a}), 176 ({\rm
  b}), 174 ({\rm c}) $, and $172 ({\rm d}) \:\:{\rm GeV}$,
respectively.}  
\label{fig:1}
\end{center}
\end{figure}

The  situation is demonstrated  in Figure \ref{fig:1}, where radiative 
corrections  \cite{1loop} 
 \begin{eqnarray}
m_{W}^{2}=\frac{1}{2}\Bigg \{
m_{Z}^{2}+\sqrt{m_{Z}^{4}-\frac{4\pi \alpha }
{\sqrt{2}G_{\mu}}m_{Z}^{2}(1+\Delta r)}
\Bigg \}, 
\label{eq:mw}
\end{eqnarray}
where 
\begin{eqnarray}
\Delta r&=&
\frac{1}{m_{W}^{2}}\left ({\rm Re}A_{WW}(m_{W}^{2})-{\rm Re} 
A_{WW}(0) \right )-\frac{2\delta e }{e}
\nonumber \\
& &
+\frac{c^{2}}{s^{2}}
\left (\frac{1}{m_{Z}^{2}}{\rm Re}A_{ZZ}(m_{Z}^{2})
-\frac{1}{m_{W}^{2}}{\rm Re} A_{WW}(m_{W}^{2}) \right )
\nonumber \\
& & +\frac{g^{2}}{16\pi ^{2}}\bigg \{
-8\left  ( \frac{1}{n-4}+C +{\rm log}\frac{m_{Z}}{\mu} \right )
%\nonumber \\
%& & \hskip2cm 
+\frac{{\rm log}\: c^{2}}{s^{2}} 
\left ( \frac{7}{2}-6s^{2} \right )+6
\bigg \}, 
\label{eq:deltar}
\end{eqnarray}
are evaluated with the input data 
\begin{eqnarray}
m_{Z}&=&91.1876\:\:\: {\rm GeV}, 
\label{eq:mz}
\\
\alpha ^{-1}&=&137.03599911, 
\label{eq:alpha}
\\
G_{\mu}&=&1.16637 \times 10^{-5}\;\;\; {\rm GeV}^{-2}, 
\label{eq:gmu}
\end{eqnarray}
for $m_{W}$ as a function of the neutral Higgs boson mass $m_{\phi}$
with the top quark mass shown as a varying parameter. In
eq. (\ref{eq:deltar}) $C=(\gamma _{E}-{\rm log}4\pi )/2$, and $s^{2}$
and $c^{2}$ stand for ${\rm sin}^{2}\theta _{W}$ and ${\rm
  cos}^{2}\theta _{W}$, respectively. $A_{ZZ}(q^{2})$ and
$A_{WW}(q^{2})$ are the $g^{\mu \nu}$ component of the gauge boson
self-energy; $\delta e $ is charge renormalisation. The Figure
\ref{fig:1} indicates that the allowed region is unnervingly
restricted if the top quark mass is as low as
(\ref{eq:topquakmass}). This presses us to consider some extension of
the minimal Higgs structure, especially in view of the Higgs search
with the LHC experiment which is the most imminent target.
%%%%%%%%%%%%%%%%%%%%%%%%%%%%%%%%%%

The most straightforward extension of the Higgs sector is to double
the Higgs doublet that would make the successful Higgs effects on the
gauge sector intact.  With two Higgs doublets $\Phi _{1}$ and $\Phi
_{2}$ having $Y=+1$, we consider the Higgs potential
\begin{eqnarray}
V&=&\mu _{1}^{2}\Phi _{1}^{\dag}\Phi _{1}+\mu _{2}^{2}
\Phi _{2}^{\dag}\Phi _{2}
%-
%\left ( \mu _{12}^{2} \Phi _{1}^{\dag}\Phi _{2}+
%\mu _{12}^{2*} \Phi _{2}^{\dag}\Phi _{1}
%\right )
+\lambda _{1}(\Phi _{1}^{\dag}\Phi _{1})^{2}+\lambda _{2}
(\Phi _{2}^{\dag}\Phi _{2})^{2}
+\lambda _{3}(\Phi _{1}^{\dag}\Phi _{1})(\Phi _{2}^{\dag}\Phi _{2})
\nonumber \\
& & 
+\lambda _{4}(\Phi _{1}^{\dag}\Phi _{2})(\Phi _{2}^{\dag}\Phi _{1})
+\frac{1}{2}\left (
\lambda _{5}(\Phi _{1}^{\dag}\Phi _{2})^{2}+\lambda _{5}^{*}(\Phi _{2}^{\dag}
\Phi _{1})^{2}
\right )
%\nonumber \\
%& &
%+\left (
%\lambda _{6}(\Phi _{1}^{\dag}\Phi _{1})(\Phi _{1}^{\dag}\Phi _{2})
%+
%\lambda _{6}^{*}(\Phi _{1}^{\dag}\Phi _{1})(\Phi _{2}^{\dag}\Phi _{1})
%\right )
%\nonumber \\
%& &
%+\left (
%\lambda _{7}(\Phi _{2}^{\dag}\Phi _{2})(\Phi _{1}^{\dag}\Phi _{2})
%+
%\lambda _{7}^{*}(\Phi _{2}^{\dag}\Phi _{2})(\Phi _{2}^{\dag}\Phi _{1})
%\right )
.
\label{eq:potential}
\end{eqnarray}
We do not include those terms as 
$(\Phi _{1}^{\dag}\Phi _{1})(\Phi _{1}^{\dag}\Phi _{2})$ 
and 
$(\Phi _{2}^{\dag}\Phi _{2})(\Phi _{1}^{\dag}\Phi _{2})$, 
imposing the discrete symmetry under $\Phi _{2} \rightarrow -\Phi _{2}$ 
for the quartic couplings, 
to avoid flavour-changing neutral currents
\cite{glashowweinberg}. 
The quadratic mixing terms $\Phi _{1}^{\dag}\Phi _{2}$ 
and its hermitian conjugate that are 
analogous to the $\mu-$term in the minimally supersymmetric model %(MSSM)
are also omitted lest our analyses should become  too clumsy.

The particle content of the two-Higgs doublet model (THDM)  
  may be seen by putting 
\begin{eqnarray}
\Phi _{1}=\left (
\begin{tabular}{c}
$w_{1}^{\dag}$
\\
$\frac{1}{\sqrt{2}}\left (
v_{1}+h_{1}+iz_{1}
\right )$
\end{tabular}
\right ),
\:\:\:
\Phi _{2}=\left (
\begin{tabular}{c}
$w_{2}^{\dag}$
\\
$\frac{1}{\sqrt{2}}\left (
v_{2}+h_{2}+iz_{2}
\right )$
\end{tabular}
\right )
\end{eqnarray}
into the Higgs potential, where the vacuum expectation values $v_{1}$ 
and $v_{2}$ 
satisfy   
%\begin{eqnarray}
$
v_{1}^{2}+v_{2}^{2}\equiv v^{2}=(246\:\:\: {\rm GeV})^{2}. 
$
%\end{eqnarray}
The CP-even neutral Higgs bosons, $h$ and $H$, 
are obtained from $h_{1}$ and $h_{2}$ by  diagonalizing the mass 
terms with an angle $\alpha$
\begin{eqnarray}
\left (
\begin{tabular}{c}
$h_{1}$
\\
$h_{2}$
\end{tabular}
\right )=\left (
\begin{tabular}{cc}
${\rm cos}\alpha $ & $-{\rm sin}\alpha $
\\
${\rm sin}\alpha $ & ${\rm cos}\alpha $
\end{tabular}
\right )
\left (
\begin{tabular}{c}
$h$
\\
$H$
\end{tabular}
\right ). 
\label{eq:mixing1}
\end{eqnarray}
Likewise, the charged Higgs boson $H^{\pm}$, the CP-odd neutral Higgs 
boson $A^{0}$ and the Nambu-Goldstone bosons $w$ and $z$ are  
given by the rotation with the angle $\beta $,  %determined by 
${\rm tan}\beta =v_{2}/v_{1}$
\begin{eqnarray}
& & \left (
\begin{tabular}{c}
$z_{1}$
\\
$z_{2}$
\end{tabular}
\right )=\left (
\begin{tabular}{cc}
${\rm cos}\beta $ & $-{\rm sin}\beta $
\\
${\rm sin}\beta $ & ${\rm cos}\beta $
\end{tabular}
\right )
\left (
\begin{tabular}{c}
$z$
\\
$A^{0}$
\end{tabular}
\right ), 
\\
& & \left (
\begin{tabular}{c}
$w_{1}$
\\
$w_{2}$
\end{tabular}
\right )=\left (
\begin{tabular}{cc}
${\rm cos}\beta $ & $-{\rm sin}\beta $
\\
${\rm sin}\beta $ & ${\rm cos}\beta $
\end{tabular}
\right )
\left (
\begin{tabular}{c}
$w$
\\
$H^{\pm}$
\end{tabular}
\right ). 
\label{eq:mixing3}
\end{eqnarray} 
The seven parameters in the original Higgs potential (\ref{eq:potential}), 
%i.e., 
$\mu _{1}$, $\mu _{2}$, $\lambda _{1}$, $\lambda _{2}$, $\lambda _{3}$, 
$\lambda _{4}$, and $\lambda _{5}$
are now replaced by the vacuum expectation value 
$v$, the mixing angles $\alpha $ and $\beta $ and the 
four Higgs masses,  $m_{h}$, $m_{H}$, $m_{H^{\pm}}$, and $m_{A^{0}}$.

The radiative corrections in THDM has already been studied in the literature
\cite{bertolini} - \cite{toussaint}. (Bertolini \cite{bertolini} has 
in fact computed the THDM contributions to $A_{WW}(q^{2})$ 
and $A_{ZZ}(q^{2})$). 
The numerical analyses of $\Delta r$  
using the formulae such as those given in \cite{bertolini}, however, 
 would be  awkward if we scrutinize every corner of 
parameter space of $m_{h}$, $m_{H}$, $m_{H^{\pm}}$, $m_{A^{0}}$, 
$\alpha $, and $\beta $, and it would be hard to grasp the
structure of the model.  
To understand the structure of the Higgs sector, we want to reduce the
parameter space.
In fact,
a considerable simplification  takes place, if we set 
\begin{eqnarray}
m_{h}=m_{H}
\label{eq:mh=mH}
\end{eqnarray}
with which the mixing angles, $\alpha $ and $\beta $ 
disappear in $A_{WW}(q^{2})$ and $A_{ZZ}(q^{2})$. 
This reduction of parameter space makes  the numerical analyses  
significantly more transparent.

Upon requiring (\ref{eq:mh=mH}),  
many  of the terms  in $A_{WW}(q^{2})$ and $A_{ZZ}(q^{2})$ 
coincide with those that appear in the standard model calculation. 
Let us denote 
the standard model Higgs contributions to $A_{WW}(q^{2})$ by 
$\delta A_{WW}(q^{2})$ and the full THDM contributions by 
$\Delta A_{WW}(q^{2})$, and  similarly $\delta A_{ZZ}(q^{2})$ 
and $\Delta A_{ZZ}(q^{2})$ 
for $A_{ZZ}(q^{2})$, respectively. 
Writing  $m_{\phi }=m_{H}$ in the standard model, 
the  relation between THDM and standard model calculations is 
\begin{eqnarray}
& & \hskip-1cm \Delta A _{WW}(q^{2})
\Bigg \vert _{m_{h}=m_{H}}
=
\delta A _{WW}(q^{2}) \Bigg \vert _{m_{\phi }=m_{H}}
+\frac{g^{2}}{16\pi ^{2}}
\left ( \frac{2}{n-4}+2C-1 \right )\times \frac{1}{6}q^{2}
\nonumber \\
& &
+\frac{g^{2}}{16\pi ^{2}}\Bigg \{
-\frac{1}{2}K_{1}(m_{H^{\pm }}^{2}, m_{A^{0}}^{2}, q^{2})-\frac{1}{2}
K_{1}(m_{H^{\pm }}^{2}, m_{H}^{2}, q^{2})+\frac{1}{4}m_{H}^{2}
{\rm log}\frac{m_{H}^{2}}{\mu ^{2}}
\nonumber \\
& &+\frac{1}{2}m_{H^{\pm}}^{2}{\rm log}\frac{m_{H^{\pm}}^{2}}{\mu ^{2}}
+\frac{1}{4}m_{A^{0}}^{2}{\rm log}\frac{m_{A^{0}}^{2}}{\mu ^{2}}
\Bigg \}, 
\label{eq:959}
\end{eqnarray}
\begin{eqnarray}
& & \hskip-1cm \Delta A _{ZZ}(q^{2})
\Bigg \vert _{m_{h}=m_{H}}
=
\delta A _{ZZ}(q^{2})
\Bigg \vert _{m_{\phi }=m_{H}}
+\frac{g^{2}}{16\pi ^{2}}
\left ( \frac{2}{n-4}+2C-1 \right )\times \frac{1+
(c^{2}-s^{2})^{2}}{12c^{2}}q^{2}
\nonumber \\
& &
+\frac{g^{2}}{16\pi ^{2}}\frac{1}{c^{2}}\Bigg \{
-\frac{1}{2}K_{1}(m_{A^{0}}^{2}, m_{H}^{2}, q^{2})
+\frac{1}{4}m_{H}^{2}{\rm log}\frac{m_{H}^{2}}{\mu ^{2}}
+\frac{1}{4}m_{A^{0}}^{2}{\rm log}\frac{m_{A^{0}}^{2}}{\mu ^{2}}
\Bigg \}
\nonumber \\
& &
+ \frac{g^{2}}{16\pi ^{2}}\frac{(c^{2}-s^{2})^{2}}{c^{2}}
\Bigg \{
-\frac{1}{2}K_{1}(m_{H^{\pm}}^{2}, m_{H^{\pm}}^{2}, q^{2})
+\frac{1}{2}m_{H^{\pm}}^{2}{\rm log}\frac{m_{H^{\pm}}^{2}}{\mu ^{2}}
\Bigg \},
\label{eq:987}
\end{eqnarray}
where 
$K_{1}$ 
%and $K_{2}$ are 
is the integral,
\begin{eqnarray}
K_{1}(m_{1}^{2}, m_{2}^{2}, q^{2})
&=&
\int _{0}^{1} dx \:\left (
m_{1}^{2}x +m_{2}^{2}(1-x)-q^{2}x(1-x)
\right )
\nonumber \\
& & \times {\rm log}\left (
\frac{m_{1}^{2}x+m_{2}^{2}(1-x)-q^{2}x(1-x)}{\mu ^{2}}\right ). 
\label{eq:kone}
%\\
%K_{2}(m_{1}^{2}, m_{2}^{2}, q^{2})
%&=&
%\int _{0}^{1} dx \:
%{\rm log}\left ( \frac{m_{1}^{2}x+m_{2}^{2}(1-x)-q^{2}
%x(1-x)}{\mu ^{2}}\right ).
%\label{eq:ktwo}
\end{eqnarray}
%%%%%%%%%%%%%%%%%%%%%%%%%%%%%%%%%%%%%%%%
%Neither (\ref{eq:959}) nor (\ref{eq:987}) depends on 
%the mixing angles $\alpha $ and  $\beta$. 

%\vfill\eject
%%%%%%%%%%%%%%%%%%%%%%%%%%%%
For electric charge renormalisation $\delta e$ 
%was also discussed  in \cite{1loop} in SM . In the case of THDM, 
we add the contribution from the charged Higgs boson % and 
to the  two-point function of the electromagnetic currents,
$\Pi _{\gamma \gamma}^{({\rm charged})}(0)$,
%in addition 
to the leptonic and hadronic contributions in the minimal model:
%$\delta e $ is expressed as 
%\begin{eqnarray}
%\frac{2\delta e }{e}=
%-\Pi _{\gamma \gamma}^{(\ell)}(0)-\Pi _{\gamma \gamma }^{(h)}(0)
%-\Pi _{\gamma \gamma}^{({\rm charged})}(0)
%-\frac{7e^{2}}{8\pi ^{2}}\left (
%\frac{1}{n-4}+C+{\rm log}\frac{m_{W}}{\mu}-\frac{1}{21}\right )
%\end{eqnarray}
%Here $\Pi _{\gamma \gamma }^{(\ell)}$, 
%$\Pi _{\gamma \gamma }^{(h)}$, and 
%$\Pi _{\gamma \gamma }^{({\rm charged})}$
%are, respectively,  the leptonic, the hadronic and the charged Higgs boson 
%contributions to the two-point function of the electromagnetic currents
%\begin{eqnarray}
%\Pi _{\gamma \gamma }^{\mu \nu}(q^{2})=A_{\gamma \gamma}(q^{2})g^{\mu \nu}
%+B_{\gamma \gamma }(q^{2})q^{\mu}q^{\nu}, \hskip1cm A_{\gamma \gamma}
%(q^{2})=-q^{2}\Pi _{\gamma \gamma}(q^{2}).
%\end{eqnarray}
%The contributions due to the charged Higgs bosons to the photon 
%propagator  are:
\begin{eqnarray}
%\Delta A _{\gamma \gamma }(q^{2})
-q^{2}\Pi _{\gamma \gamma}^{{(\rm charged)}}(q^{2})
%
%&=&
%A_{\gamma \gamma }(q^{2})\bigg \vert _{\rm THDM\:\:Higgs}
%\nonumber \\
%&=&-q^{2}\Pi _{\gamma \gamma}^{{(\rm charged)}}(q^{2})
%\nonumber \\
&=&+\frac{g^{2}}{16\pi ^{2}}g^{\mu \nu} \: s^{2}\: 
\left ( \frac{2}{n-4}+2C-1 \right ) \times \frac{1}{3}q^{2}
\nonumber \\
& &+ \frac{g^{2}}{16\pi ^{2}}\: s^{2} \: \Bigg \{
-2K_{1}(m_{H^{\pm}}^{2}, m_{H^{\pm}}^{2}, q^{2})
+2m_{H^{\pm}}^{2}{\rm log}\frac{m_{H^{\pm}}}{\mu ^{2}}
\Bigg \}.
\label{eq:thdmgg}
\end{eqnarray}

We write $\Delta r$ in two parts:
\begin{eqnarray}
\Delta r=\Delta r^{(1)}+ \Delta r ^{(2)}
\end{eqnarray}
where 
$\Delta r ^{(1)}$ is the term that appears 
in the standard model and 
$\Delta r^{(2)}$ is the contribution from 
the charged  and CP-odd Higgs bosons: 
\begin{eqnarray}
& & \hskip-1cm 
\Delta r^{(2)}=\frac{g^{2}}{16\pi ^{2}}\frac{1}{m_{W}^{2}}\Bigg  \{
\left ( \frac{c^{2}}{2s^{2}} -\frac{1}{2} \right )
\left [
K_{1}(m_{H^{\pm}}^{2}, m_{A^{0}}^{2}, m_{W}^{2})+
K_{1}(m_{H^{\pm}}^{2}, m_{H}^{2}, m_{W}^{2})
\right  ] 
\nonumber \\
& &
-\frac{c^{2}}{2s^{2}}K_{1}(m_{A^{0}}^{2}, m_{H}^{2}, m_{Z}^{2})
+\frac{1}{2}K_{1}(m_{H^{\pm}}^{2}, m_{A^{0}}^{2}, 0)
+\frac{1}{2}K_{1}(m_{H^{\pm}}^{2}, m_{H}^{2}, 0)
\nonumber \\
& &
-\frac{c^{2}}{2s^{2}}m_{H^{\pm}}^{2}{\rm log}\frac{m_{H^{\pm}}^{2}}{\mu ^{2}}
\Bigg \}
\nonumber \\
& &
+\frac{g^{2}}{16\pi ^{2}}\frac{(c^{2}-s^{2})^{2}}{s^{2}}\frac{1}{m_{Z}^{2}}
\Bigg \{
-\frac{1}{2}K_{1}(m_{H^{\pm}}^{2}, m_{H^{\pm}}^{2}, m_{Z}^{2})
+\frac{1}{2}m_{H^{\pm}}^{2}{\rm log}\frac{m_{H^{\pm}}^{2}}{\mu ^{2}}
\Bigg \}
\nonumber \\
& &
+\frac{g^{2}}{16\pi ^{2}}\left (
1+{\rm log}\frac{m_{H^{\pm}}^{2}}{\mu ^{2}}
\right )\times \left ( -\frac{1}{3}s^{2} \right )\ .
\label{eq:deltar2}
\end{eqnarray}
%As we see clearly, 
%$\Delta r^{(2)} $ depends  only on two 
%parameters i.e., $m_{H^{\pm}}$ and $m_{A^{0}}$.

%%%%%%%%%%%%%%%%%%%%%%%%%%%%%%%%%%%%%%
%\vfill\eject
%\section{Numerical Evaluation}\label{}
%\setcounter{equation}{0}
%\setcounter{figure}{0}
%\renewcommand{\theequation}{7.\arabic{equation}}
%\renewcommand{\thefigure}{7.\arabic{figure}}

In our numerical exploration  in what follows we 
consider only the case where the CP-even Higgs bosons, $h$ and $H$,  
are degenerate.
There are then no mixing parameters $\alpha $ and $\beta $ in
(\ref{eq:deltar2}) appearing, leaving us with only three parameters,
$m_H$, $m_{H^{\pm}}$, and  $m_{A^{0}}$. 
Note that the Higgs-gauge coupling depends only on the 
difference $\alpha - \beta$ from the 
structure of the mixing (\ref{eq:mixing1}) - (\ref{eq:mixing3}) and 
the dependence on $\alpha $ and $\beta $ disappears simultaneously by 
setting (\ref{eq:mh=mH}).

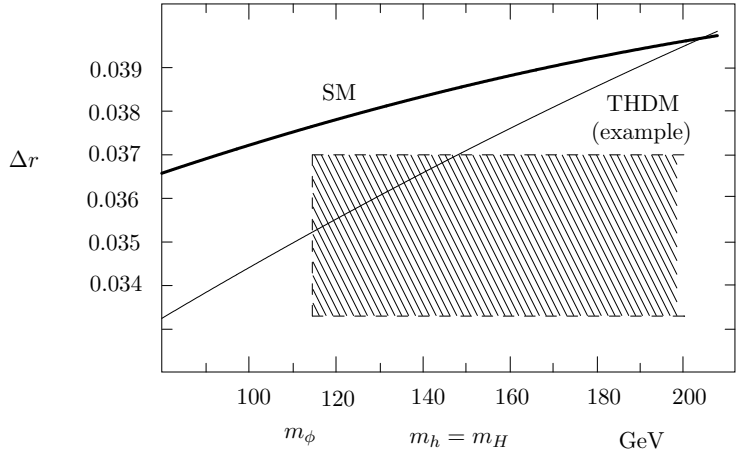
\begin{figure}[b]
\begin{center}
%\includegraphics{mhdr.eps}
%\hskip1cm 
%\special{epsfile=fig2.eps hscale=0.85, vscale=0.85}
\scalebox{0.8}[0.8]{
\input{fig2X.tex}}
%\vskip7cm
\caption{
Radiative correction to the $W$ boson mass, $\Delta r$,
in the on-shell renormalisation scheme for the standard model
(SM) with minimal Higgs sector (thick line) and for an example of
the THDM (thin line), where $m_{A^{0}}=150$ GeV and 
$m_{H^{\pm}}=200 $ GeV are adopted. The top quark mass is assumed to be 
172 GeV. 
}  
\label{fig:2}
\end{center}
\end{figure}

Figure \ref{fig:2} shows the correction $\Delta r$ for the minimal model
and an example of the THDM as a function of the neutral Higgs boson mass.  The
empirically viable region is indicated by shading.  For the THDM
radiative correction we take, as an example, $m_{H^\pm}=200$ GeV and
$m_{A^0}=150$ GeV. The figure shows that the contribtions of the
charged and/or CP odd neutral Higgs bosons cancel partly the correction from
the CP even neutral Higgs bosons, making the radiative
correction consistent with the empirical $W$ boson mass even with a
large CP even neutral Higgs mass that would be unfavoured in the
minimal Higgs model. 
%This leads to a constraint on the charged and CP
%odd neutral Higgs boson mass for given CP even neutral Higgs mass.

\begin{figure}[b]
\begin{center}
\scalebox{0.65}[0.65]{
\input{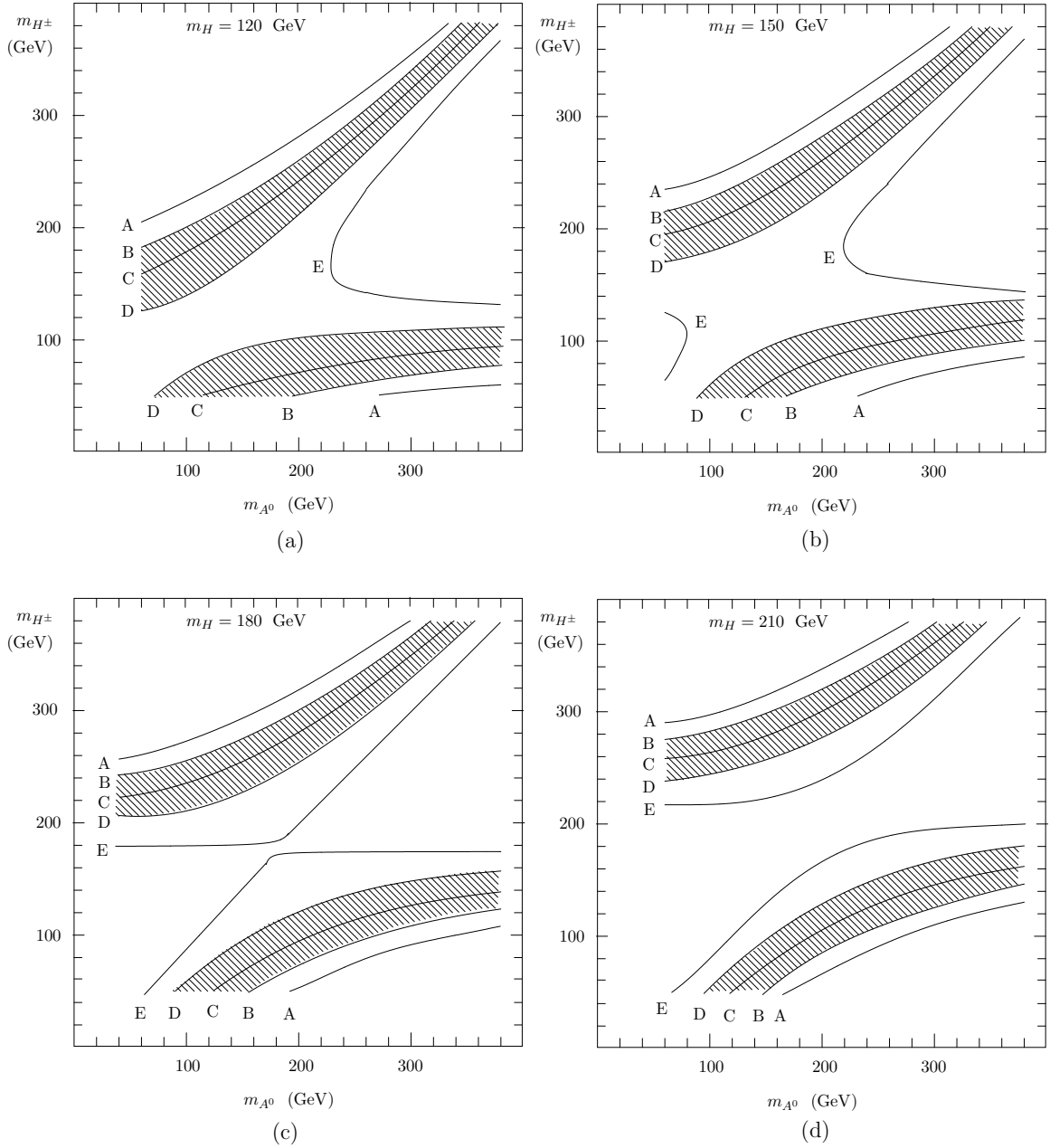}}
\caption{
The region expected from the radiative correction 
for the $(m_{A^{0}}, m_{H^{\pm}})$ plane when 
(a) $m_{H}=120$ GeV, (b) 150 GeV, (c) 180 GeV and (d) 210 GeV.
We fix top quark mass at $m_{t}=172$ GeV.
The curves correspond to 
(A) $m_{W}=80.46$, (B) 80.43, (C) 80.40, (D), 80.37, and (E) 80.34 GeV.
The region empirically allowed by the $W$ mass is shown by shading.
%
%Radiative correction to the $W$ boson mass $\Delta r$
%in the on-shell renormalisation scheme for the standard model
%with minimal Higgs sector (thick line) and for an example of
%the THDM (thin line). For the THDM, ..... are taken. 
}
\label{fig:3}
\end{center}
\end{figure}

Figure \ref{fig:3}  shows the region favoured for the charged and CP
odd neutral Higgs boson masses for a given CP even neutral Higgs mass,
which is taken to be 120, 150, 180 and 210 GeV. We see that there are
two separate regions for any $m_H$ allowed for $(m_{H^\pm},m_{A^0})$, one 
along the oblique line $m_{H^\pm}\approx m_{A^0}$ for large mass 
and the other in a lower plane 
with $m_{H^\pm}$ increasing only  little with $m_{A^0}$.
The four panels (a)$-$(d) show that the allowed regions are generally 
shifted upwards as
$m_H$ increases, but the increments depend on specific branches.  
The smallest mass
of the upper branch with respect to $m_H$ is nearly constant, 
$m_{H^\pm}/m_H\approx 1.1-1.2$. For the large mass region of the same branch
the upward shift with $m_H$ is modest, and the asymptotics seem converging
to $m_{H^\pm}= m_{A^0}$.
The asymptotics of the lower branch also increse with $m_H$, but
the solution includes the charged Higgs mass staying at 60 $-$ 180 GeV, 
irrespective of
the neutral Higgs boson masses. There is always a solution with
$m_{H^\pm}<100$ GeV, which would yield the required radiative correction.
We emphasise that the Higgs mass ratios $m_{H^\pm}/m_H$ and
$m_{A^0}/m_H$ would give a useful indicator of the THDM for the Higgs
sector through the electroweak radiative correction.

%%%%%%%%%%%%%%%%%%%%%%%%%%%%%%%%%%%%%%%%%%%%%%%%%%%%%%%%%%%%%%%%%%%%%
%\vfill\eject
%\begin{center}
%\begin{figure}[htb]
%\begin{center}
%\includegraphics{mhdr.eps}
%\hskip1cm 
%\special{epsfile=180GeVcontour5.eps hscale=0.8, vscale=0.8}
%\special{epsfile=guido.eps hscale=1.00, vscale=1.00}
%\vskip11cm
%\caption{
%The region expected from the radiative correction 
%for the $(m_{A^{0}}, m_{H^{\pm}})$ plane when 
%(a) $m_{H}=120$ GeV, (b) 150 GeV, (c) 180 GeV and (d) 210 GeV.
%We fix top quark mass at $m_{t}=172$ GeV.
%The curves corerespond to 
%(a) $m_{W}=80.46$, (b) 80.43, (c) 80.40, (d), 80.37, and (e) 80.34 GeV.
%The region empirically allowed by the $W$ mass is shown by shading.
%}
%\end{center}
%\label{fig:180}
%\end{figure}
%\end{center}
%%%%%%%%%%%%%%%%%%%%%%%%%%%%%%%%%%%%%%%%%%%%%%%%%%%%%%%%%%%%%%%%%%%%%

The regions favoured by radiative correctons can be seen
in a simple argument. 
Noting that for large $M$, $K_1$ behaves as
\begin{eqnarray}
& & \lim _{M\to \infty}K_{1}(M^{2}, M^{2}, q^{2})
\sim 
\frac{1}{2}M^{2}{\rm log}\left ( \frac{M^{2}}{\mu ^{2}}\right )
-\frac{1}{4}M^{2}
+ {\cal O}\left ( q^{2}  \right ), 
\label{eq:asymptot1}
\\
& & \lim _{M\to \infty} K_{1}(M^{2}, m^{2}, q^{2})
\sim 
M^{2}{\rm log}\left (\frac{M^{2}}{\mu ^{2}}
\right )
+{\cal O}\left ( q^2, m^{2} \right ), 
\label{eq:asymptot2}
\end{eqnarray}
%\\
we find for the limiting case $m_{H^{\pm}}=m_{A^{0}}=M \to \infty$
along $m_{H^{\pm}}\approx m_{A^{0}}$,
\begin{eqnarray}
& & 
%\lim _{M\to \infty} 
\Delta A_{WW}(q^{2})
\vert _{m_{h}=m_{H}}
\sim \frac{g^{2}}{16\pi ^{2}}
\times \frac{1}{8}M^{2}, 
\hskip0.5cm 
( {\rm as } \:\:\: m_{H^{\pm}}=m_{A^{0}}=M\to \infty ),  
\label{eq:large1}
\\
& & 
%\lim _{M\to \infty} 
\Delta A_{ZZ}(q^{2})
\vert _{m_{h}=m_{H}}
\sim \frac{g^{2}}{16\pi ^{2}}
\times \frac{1}{8 c^{2}}M^{2}, 
\hskip0.5cm
({\rm as} \:\:\: m_{H^{\pm}}=m_{A^{0}}=M\to \infty ),
\label{eq:large2}
\end{eqnarray}
%This type of non-decopuling of the Higgs contribution was noticed 
%long time ago in \cite{toussaint} and \cite{bertolini}. 
which lead to a cancellation in 
${\rm Re}A_{WW}(m_W^2) - {\rm Re}A_{WW}(0)$ and also in 
$m_Z^{-2}{\rm Re}A_{ZZ}(m_Z^2) - m_W^{-2}{\rm Re}A_{WW}(m_W^2)$
in the expression of
$\Delta r$.
This means that allowed regions exist even for a very large $m_H$
in the vicinity of the line $m_{H^{\pm}}=m_{A^{0}}$.

We also see a similar cancellation in the limit
$m_{A^0}\rightarrow\infty$ with $m_H$ and $m_{H^{\pm}}$ fixed.
We obtain
\begin{eqnarray}
\Delta A_{WW}(q^2)\Bigg \vert _{m_{h}=m_{H}}\sim 
\frac{g^2}{16\pi ^2}\times \left ( -\frac{1}{4}\right )M^2 {\rm log}
\left ( \frac{M^2}{\mu ^2}\right ), 
\:\:\: ({\rm as} \:\:\: m_{A^{0}}=M \to \infty)
\label{eq:772}
\\
\Delta A_{ZZ}(q^2)\Bigg \vert _{m_{h}=m_{H}}\sim 
\frac{g^2}{16\pi ^2}\times \left ( -\frac{1}{4c^2}\right )M^2 {\rm log}
\left ( \frac{M^2}{\mu ^2}\right ). 
\:\:\: ({\rm as} \:\:\: m_{A^{0}}=M \to \infty)
\label{eq:778}
\end{eqnarray}
This causes a similar cancellation in $\Delta r$, 
making the second region where
$m_{H^{\pm}}\approx$ constant allowed in the figure.

\begin{figure}[htb]
\begin{center}
%\includegraphics{mhdr.eps}
%\hskip1cm 
%\special{epsfile=ma150.eps hscale=0.8, vscale=0.8}
%\special{epsfile=guido.eps hscale=1.00, vscale=1.00}
%\vskip11cm
\scalebox{0.8}[0.8]{
\input{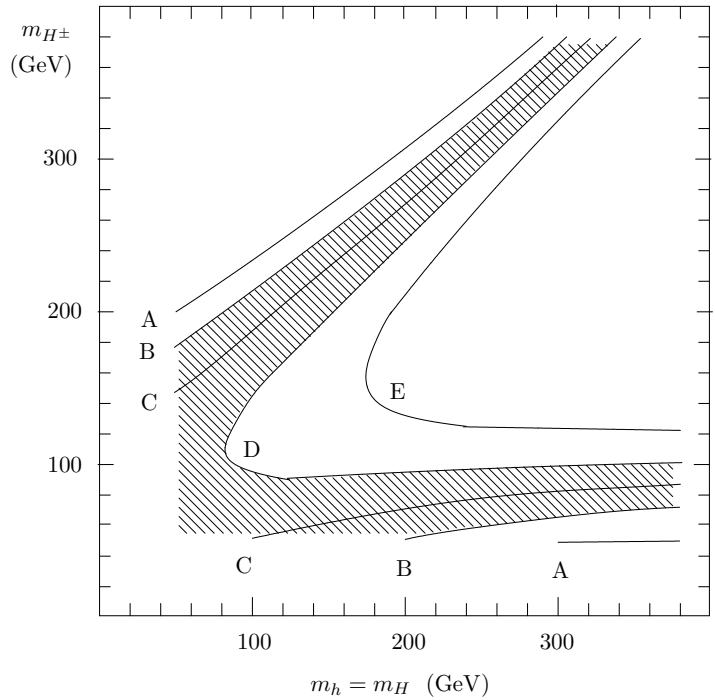}}
\caption{Region favoured from the radiative correction for $m_H^\pm$
as a function of $m_H$ in the THDM. The CP odd neutral Higgs boson
is taken to have $m_{A^0}=150$ GeV. The curves correspond to
(A) $m_W=80.46$, (B) $80.43$,(C) $80.40$, 
(D) $80.37$, (E) $80.34$ GeV,
and the region with hatched is allowed. 
We fix top quark mass at $m_{t}=172$ GeV.
}
\label{fig:4}
\end{center}
\end{figure}

%%%%%%%%%%%%%%%%%%%%%%%%%%%%%%%%%%%%%%%%%%%%%%%%%%%%%%%%%%%%%%%%%%%%%
We also study the case for both $m_H$ and $m_{H^{\pm}}$ being large.
The limiting case can also be seen with a
limiting case 
similar to the above,
$m_{H^{\pm}}=m_{H}=M\to \infty$. We then arrive at expressions for
$\Delta A_{WW}(q^{2})$ and $\Delta A_{ZZ}(q^{2})$ the same as
eqs. (\ref{eq:772}) and (\ref{eq:778}) but merely by a factor 
of 2 larger. This shows a cancellation
of the radiative corrections, which means the presence of a preferred region
in the vicinity of the line in agreement with the conclusion we derived
for Figure \ref{fig:3} . In a similar way we see a preferred region along
$m_{H^{\pm}}=$const.
We show the preferred region of 
$m_{H^{\pm}}$ that is correlated with $m_H$ in 
Figure \ref{fig:4} for a specified $m_{A^0}$.

In the present paper we show that the electroweak radiative correction,
which is marginally consistent with the present experiment in the
minimal Higgs model in that the Higgs mass preferred by the radiative
correction appears too low compared with the lower limit from the Higgs
search, is well relaxed if the Higgs sector is endowed with
two Higgs doublets. We explored the mass region where the charged
and CP odd neutral Higgs bosons may reside, albeit in a restricted
model where two CP even Higgs bosons are degenerate in mass, which
made the systematic exploration feasible. The mass ratios among
Higgs bosons to be looked for at LHC would serve as useful indicators
to explore the Higgs sector via electroweak radiative corrections.

\vfill\eject
\vskip1cm
\noindent
{\large {\bf Acknowledgement}}

One of us (T.K.) thanks Luis Alvarez-Gaume 
for his hospitality at CERN Theory Division. 
MF is supported by Grant in Aid of the Ministry of Education.

%\vfill\eject
\vskip1cm

\vfill\eject
%%%%    figure  1  %%%%%%%%%%%%%%%%%%%%%%%%%%%%%%
%\begin{center}
%\end{center}
%%%%%%%%%%%%%%%%%%%%%%%%%%%%%%%%%%%%

\vfill\eject
%%%%%%%%%%%%%%   figure 2   %%%%%%%%%%%%%%%%%%%%

%%%%%%%%%%%%%%%%%%%%%%%%%%%%%%%%%%%%

%%%%%%%%%%%%%%%%%%%    figure 3   %%%%%%%%%%%%%%%%%%%%
\vfill\eject
\begin{figure}[t]
\begin{center}
%\includegraphics{mhdr.eps}
%\hskip1cm 
%\special{epsfile=fig3A.eps hscale=0.8, vscale=0.8}
%\hskip8cm
%\special{epsfile=fig3B.eps hscale=0.8, vscale=0.8}
%\vskip8cm
%\special{epsfile=fig3C.eps hscale=0.8, vscale=0.8}
%\hskip8cm
%\special{epsfile=fig3D.eps hscale=0.8, vscale=0.8}
%\special{epsfile=guido.eps hscale=1.00, vscale=1.00}
%\vskip7cm
\scalebox{0.65}[0.65]{
\input{fig3ABX.tex}}
\end{center}
\end{figure}

%%%%%%%%%%%%%%%%%%%%%%%%%%%%%%%%%%%%%%%%%%%%%%%%%%
%\vfill\eject
%%%%%%%%%%%%%%%%%%%%%%%%%%%%%%%%%%%%%%%%%%%%%%%

%%%%%%%%%%%%%%%%%%%%%%%%%%%%%%%%%%%%%%%%%%%%%%%%%%%%%%%%%%%%%%%%%%%
\vfill\eject

%%%%%%%%%%%%%%%%%%%%%%%%%%%%%%%%%%%%%%%%%%%%%%%

\end{document}

%% file: fig2X.tex
%WinTpicVersion2.15
\unitlength 0.1in
\begin{picture}(47.50,27.80)(6.40,-33.00)
% STR 2 0 3 0
% 3 1170 2768 1170 2840 33 206
% 0.034
\put(11.7000,-24.4000){\makebox(0,0)[lb]{0.034}}%
% STR 2 0 3 0
% 3 1160 2499 1160 2570 33 206
% 0.035
\put(11.6000,-21.7000){\makebox(0,0)[lb]{0.035}}%
% STR 2 0 3 0
% 3 1150 2218 1150 2290 33 206
% 0.036
\put(11.5000,-18.9000){\makebox(0,0)[lb]{0.036}}%
% STR 2 0 3 0
% 3 1160 1919 1160 1990 33 206
% 0.037
\put(11.6000,-15.9000){\makebox(0,0)[lb]{0.037}}%
% STR 2 0 3 0
% 3 1160 1648 1160 1720 33 206
% 0.038
\put(11.6000,-13.2000){\makebox(0,0)[lb]{0.038}}%
% LINE 2 0 3 0
% 2 1644 1377 1708 1377
% 
\special{pn 8}%
\special{pa 1644 977}%
\special{pa 1708 977}%
\special{fp}%
% LINE 2 0 3 0
% 2 1644 1662 1708 1662
% 
\special{pn 8}%
\special{pa 1644 1262}%
\special{pa 1708 1262}%
\special{fp}%
% LINE 2 0 3 0
% 2 1644 1948 1708 1948
% 
\special{pn 8}%
\special{pa 1644 1548}%
\special{pa 1708 1548}%
\special{fp}%
% LINE 2 0 3 0
% 2 1644 2233 1708 2233
% 
\special{pn 8}%
\special{pa 1644 1833}%
\special{pa 1708 1833}%
\special{fp}%
% LINE 2 0 3 0
% 2 1644 2519 1708 2519
% 
\special{pn 8}%
\special{pa 1644 2119}%
\special{pa 1708 2119}%
\special{fp}%
% LINE 2 0 3 0
% 2 1644 2804 1708 2804
% 
\special{pn 8}%
\special{pa 1644 2404}%
\special{pa 1708 2404}%
\special{fp}%
% LINE 2 0 3 0
% 2 1644 3090 1708 3090
% 
\special{pn 8}%
\special{pa 1644 2690}%
\special{pa 1708 2690}%
\special{fp}%
% STR 2 0 3 0
% 3 2130 3508 2130 3580 33 206
% 100
\put(21.3000,-31.8000){\makebox(0,0)[lb]{100}}%
% STR 2 0 3 0
% 3 2680 3519 2680 3590 33 206
% 120
\put(26.8000,-31.9000){\makebox(0,0)[lb]{120}}%
% STR 2 0 3 0
% 3 3270 3509 3270 3580 33 206
% 140
\put(32.7000,-31.8000){\makebox(0,0)[lb]{140}}%
% STR 2 0 3 0
% 3 3820 3509 3820 3580 33 206
% 160
\put(38.2000,-31.8000){\makebox(0,0)[lb]{160}}%
% STR 2 0 3 0
% 3 4420 3509 4420 3580 33 206
% 180
\put(44.2000,-31.8000){\makebox(0,0)[lb]{180}}%
% STR 2 0 3 0
% 3 4980 3509 4980 3580 33 206
% 200
\put(49.8000,-31.8000){\makebox(0,0)[lb]{200}}%
% STR 2 0 3 0
% 3 1170 1369 1170 1440 33 206
% 0.039
\put(11.7000,-10.4000){\makebox(0,0)[lb]{0.039}}%
% POLYLINE 2 1 3 0
% 5 5056 1948 2625 1948 2625 3004 5056 3004 5056 3004
% 
\special{pn 8}%
\special{pa 5056 1548}%
\special{pa 2625 1548}%
\special{pa 2625 2604}%
\special{pa 5056 2604}%
\special{pa 5056 2604}%
\special{da 0.070}%
% FUNC 0 0 3 0
% 9 1644 920 5277 3090 1644 3090 5056 3090 1644 1377 1644 920 5275 3090 0 3 0 0
% (0.0362415-0.033+0.000347336*(12*x+1)-0.0000068092*(12*x+1)^2)/0.006
\special{pn 20}%
\special{pa 1640 1669}%
\special{pa 1645 1667}%
\special{pa 1650 1665}%
\special{pa 1655 1664}%
\special{pa 1660 1662}%
\special{pa 1665 1660}%
\special{pa 1670 1659}%
\special{pa 1675 1657}%
\special{pa 1680 1655}%
\special{pa 1685 1654}%
\special{pa 1690 1652}%
\special{pa 1695 1650}%
\special{pa 1700 1649}%
\special{pa 1705 1647}%
\special{pa 1710 1645}%
\special{pa 1715 1644}%
\special{pa 1720 1642}%
\special{pa 1725 1640}%
\special{pa 1730 1639}%
\special{pa 1735 1637}%
\special{pa 1740 1635}%
\special{pa 1745 1634}%
\special{pa 1750 1632}%
\special{pa 1755 1630}%
\special{pa 1760 1629}%
\special{pa 1765 1627}%
\special{pa 1770 1625}%
\special{pa 1775 1624}%
\special{pa 1780 1622}%
\special{pa 1785 1621}%
\special{pa 1790 1619}%
\special{pa 1795 1617}%
\special{pa 1800 1616}%
\special{pa 1805 1614}%
\special{pa 1810 1612}%
\special{pa 1815 1611}%
\special{pa 1820 1609}%
\special{pa 1825 1607}%
\special{pa 1830 1606}%
\special{pa 1835 1604}%
\special{pa 1840 1603}%
\special{pa 1845 1601}%
\special{pa 1850 1599}%
\special{pa 1855 1598}%
\special{pa 1860 1596}%
\special{pa 1865 1594}%
\special{pa 1870 1593}%
\special{pa 1875 1591}%
\special{pa 1880 1590}%
\special{pa 1885 1588}%
\special{pa 1890 1586}%
\special{pa 1895 1585}%
\special{pa 1900 1583}%
\special{pa 1905 1582}%
\special{pa 1910 1580}%
\special{pa 1915 1578}%
\special{pa 1920 1577}%
\special{pa 1925 1575}%
\special{pa 1930 1573}%
\special{pa 1935 1572}%
\special{pa 1940 1570}%
\special{pa 1945 1569}%
\special{pa 1950 1567}%
\special{pa 1955 1565}%
\special{pa 1960 1564}%
\special{pa 1965 1562}%
\special{pa 1970 1561}%
\special{pa 1975 1559}%
\special{pa 1980 1557}%
\special{pa 1985 1556}%
\special{pa 1990 1554}%
\special{pa 1995 1553}%
\special{pa 2000 1551}%
\special{pa 2005 1549}%
\special{pa 2010 1548}%
\special{pa 2015 1546}%
\special{pa 2020 1545}%
\special{pa 2025 1543}%
\special{pa 2030 1542}%
\special{pa 2035 1540}%
\special{pa 2040 1538}%
\special{pa 2045 1537}%
\special{pa 2050 1535}%
\special{pa 2055 1534}%
\special{pa 2060 1532}%
\special{pa 2065 1531}%
\special{pa 2070 1529}%
\special{pa 2075 1527}%
\special{pa 2080 1526}%
\special{pa 2085 1524}%
\special{pa 2090 1523}%
\special{pa 2095 1521}%
\special{pa 2100 1520}%
\special{pa 2105 1518}%
\special{pa 2110 1516}%
\special{pa 2115 1515}%
\special{pa 2120 1513}%
\special{pa 2125 1512}%
\special{pa 2130 1510}%
\special{sp}%
\special{pa 2130 1510}%
\special{pa 2135 1509}%
\special{pa 2140 1507}%
\special{pa 2145 1505}%
\special{pa 2150 1504}%
\special{pa 2155 1502}%
\special{pa 2160 1501}%
\special{pa 2165 1499}%
\special{pa 2170 1498}%
\special{pa 2175 1496}%
\special{pa 2180 1495}%
\special{pa 2185 1493}%
\special{pa 2190 1492}%
\special{pa 2195 1490}%
\special{pa 2200 1488}%
\special{pa 2205 1487}%
\special{pa 2210 1485}%
\special{pa 2215 1484}%
\special{pa 2220 1482}%
\special{pa 2225 1481}%
\special{pa 2230 1479}%
\special{pa 2235 1478}%
\special{pa 2240 1476}%
\special{pa 2245 1475}%
\special{pa 2250 1473}%
\special{pa 2255 1472}%
\special{pa 2260 1470}%
\special{pa 2265 1469}%
\special{pa 2270 1467}%
\special{pa 2275 1465}%
\special{pa 2280 1464}%
\special{pa 2285 1462}%
\special{pa 2290 1461}%
\special{pa 2295 1459}%
\special{pa 2300 1458}%
\special{pa 2305 1456}%
\special{pa 2310 1455}%
\special{pa 2315 1453}%
\special{pa 2320 1452}%
\special{pa 2325 1450}%
\special{pa 2330 1449}%
\special{pa 2335 1447}%
\special{pa 2340 1446}%
\special{pa 2345 1444}%
\special{pa 2350 1443}%
\special{pa 2355 1441}%
\special{pa 2360 1440}%
\special{pa 2365 1438}%
\special{pa 2370 1437}%
\special{pa 2375 1435}%
\special{pa 2380 1434}%
\special{pa 2385 1432}%
\special{pa 2390 1431}%
\special{pa 2395 1429}%
\special{pa 2400 1428}%
\special{pa 2405 1426}%
\special{pa 2410 1425}%
\special{pa 2415 1423}%
\special{pa 2420 1422}%
\special{pa 2425 1420}%
\special{pa 2430 1419}%
\special{pa 2435 1417}%
\special{pa 2440 1416}%
\special{pa 2445 1414}%
\special{pa 2450 1413}%
\special{pa 2455 1411}%
\special{pa 2460 1410}%
\special{pa 2465 1408}%
\special{pa 2470 1407}%
\special{pa 2475 1405}%
\special{pa 2480 1404}%
\special{pa 2485 1403}%
\special{pa 2490 1401}%
\special{pa 2495 1400}%
\special{pa 2500 1398}%
\special{pa 2505 1397}%
\special{pa 2510 1395}%
\special{pa 2515 1394}%
\special{pa 2520 1392}%
\special{pa 2525 1391}%
\special{pa 2530 1389}%
\special{pa 2535 1388}%
\special{pa 2540 1386}%
\special{pa 2545 1385}%
\special{pa 2550 1383}%
\special{pa 2555 1382}%
\special{pa 2560 1381}%
\special{pa 2565 1379}%
\special{pa 2570 1378}%
\special{pa 2575 1376}%
\special{pa 2580 1375}%
\special{pa 2585 1373}%
\special{pa 2590 1372}%
\special{pa 2595 1370}%
\special{pa 2600 1369}%
\special{pa 2605 1368}%
\special{pa 2610 1366}%
\special{pa 2615 1365}%
\special{pa 2620 1363}%
\special{sp}%
\special{pa 2620 1363}%
\special{pa 2625 1362}%
\special{pa 2630 1360}%
\special{pa 2635 1359}%
\special{pa 2640 1357}%
\special{pa 2645 1356}%
\special{pa 2650 1355}%
\special{pa 2655 1353}%
\special{pa 2660 1352}%
\special{pa 2665 1350}%
\special{pa 2670 1349}%
\special{pa 2675 1347}%
\special{pa 2680 1346}%
\special{pa 2685 1345}%
\special{pa 2690 1343}%
\special{pa 2695 1342}%
\special{pa 2700 1340}%
\special{pa 2705 1339}%
\special{pa 2710 1337}%
\special{pa 2715 1336}%
\special{pa 2720 1335}%
\special{pa 2725 1333}%
\special{pa 2730 1332}%
\special{pa 2735 1330}%
\special{pa 2740 1329}%
\special{pa 2745 1328}%
\special{pa 2750 1326}%
\special{pa 2755 1325}%
\special{pa 2760 1323}%
\special{pa 2765 1322}%
\special{pa 2770 1321}%
\special{pa 2775 1319}%
\special{pa 2780 1318}%
\special{pa 2785 1316}%
\special{pa 2790 1315}%
\special{pa 2795 1314}%
\special{pa 2800 1312}%
\special{pa 2805 1311}%
\special{pa 2810 1309}%
\special{pa 2815 1308}%
\special{pa 2820 1307}%
\special{pa 2825 1305}%
\special{pa 2830 1304}%
\special{pa 2835 1302}%
\special{pa 2840 1301}%
\special{pa 2845 1300}%
\special{pa 2850 1298}%
\special{pa 2855 1297}%
\special{pa 2860 1295}%
\special{pa 2865 1294}%
\special{pa 2870 1293}%
\special{pa 2875 1291}%
\special{pa 2880 1290}%
\special{pa 2885 1289}%
\special{pa 2890 1287}%
\special{pa 2895 1286}%
\special{pa 2900 1284}%
\special{pa 2905 1283}%
\special{pa 2910 1282}%
\special{pa 2915 1280}%
\special{pa 2920 1279}%
\special{pa 2925 1278}%
\special{pa 2930 1276}%
\special{pa 2935 1275}%
\special{pa 2940 1273}%
\special{pa 2945 1272}%
\special{pa 2950 1271}%
\special{pa 2955 1269}%
\special{pa 2960 1268}%
\special{pa 2965 1267}%
\special{pa 2970 1265}%
\special{pa 2975 1264}%
\special{pa 2980 1263}%
\special{pa 2985 1261}%
\special{pa 2990 1260}%
\special{pa 2995 1259}%
\special{pa 3000 1257}%
\special{pa 3005 1256}%
\special{pa 3010 1254}%
\special{pa 3015 1253}%
\special{pa 3020 1252}%
\special{pa 3025 1250}%
\special{pa 3030 1249}%
\special{pa 3035 1248}%
\special{pa 3040 1246}%
\special{pa 3045 1245}%
\special{pa 3050 1244}%
\special{pa 3055 1242}%
\special{pa 3060 1241}%
\special{pa 3065 1240}%
\special{pa 3070 1238}%
\special{pa 3075 1237}%
\special{pa 3080 1236}%
\special{pa 3085 1234}%
\special{pa 3090 1233}%
\special{pa 3095 1232}%
\special{pa 3100 1230}%
\special{pa 3105 1229}%
\special{pa 3110 1228}%
\special{sp}%
\special{pa 3110 1228}%
\special{pa 3115 1226}%
\special{pa 3120 1225}%
\special{pa 3125 1224}%
\special{pa 3130 1222}%
\special{pa 3135 1221}%
\special{pa 3140 1220}%
\special{pa 3145 1219}%
\special{pa 3150 1217}%
\special{pa 3155 1216}%
\special{pa 3160 1215}%
\special{pa 3165 1213}%
\special{pa 3170 1212}%
\special{pa 3175 1211}%
\special{pa 3180 1209}%
\special{pa 3185 1208}%
\special{pa 3190 1207}%
\special{pa 3195 1205}%
\special{pa 3200 1204}%
\special{pa 3205 1203}%
\special{pa 3210 1202}%
\special{pa 3215 1200}%
\special{pa 3220 1199}%
\special{pa 3225 1198}%
\special{pa 3230 1196}%
\special{pa 3235 1195}%
\special{pa 3240 1194}%
\special{pa 3245 1192}%
\special{pa 3250 1191}%
\special{pa 3255 1190}%
\special{pa 3260 1189}%
\special{pa 3265 1187}%
\special{pa 3270 1186}%
\special{pa 3275 1185}%
\special{pa 3280 1183}%
\special{pa 3285 1182}%
\special{pa 3290 1181}%
\special{pa 3295 1180}%
\special{pa 3300 1178}%
\special{pa 3305 1177}%
\special{pa 3310 1176}%
\special{pa 3315 1175}%
\special{pa 3320 1173}%
\special{pa 3325 1172}%
\special{pa 3330 1171}%
\special{pa 3335 1169}%
\special{pa 3340 1168}%
\special{pa 3345 1167}%
\special{pa 3350 1166}%
\special{pa 3355 1164}%
\special{pa 3360 1163}%
\special{pa 3365 1162}%
\special{pa 3370 1161}%
\special{pa 3375 1159}%
\special{pa 3380 1158}%
\special{pa 3385 1157}%
\special{pa 3390 1156}%
\special{pa 3395 1154}%
\special{pa 3400 1153}%
\special{pa 3405 1152}%
\special{pa 3410 1151}%
\special{pa 3415 1149}%
\special{pa 3420 1148}%
\special{pa 3425 1147}%
\special{pa 3430 1146}%
\special{pa 3435 1144}%
\special{pa 3440 1143}%
\special{pa 3445 1142}%
\special{pa 3450 1141}%
\special{pa 3455 1139}%
\special{pa 3460 1138}%
\special{pa 3465 1137}%
\special{pa 3470 1136}%
\special{pa 3475 1134}%
\special{pa 3480 1133}%
\special{pa 3485 1132}%
\special{pa 3490 1131}%
\special{pa 3495 1129}%
\special{pa 3500 1128}%
\special{pa 3505 1127}%
\special{pa 3510 1126}%
\special{pa 3515 1125}%
\special{pa 3520 1123}%
\special{pa 3525 1122}%
\special{pa 3530 1121}%
\special{pa 3535 1120}%
\special{pa 3540 1118}%
\special{pa 3545 1117}%
\special{pa 3550 1116}%
\special{pa 3555 1115}%
\special{pa 3560 1114}%
\special{pa 3565 1112}%
\special{pa 3570 1111}%
\special{pa 3575 1110}%
\special{pa 3580 1109}%
\special{pa 3585 1108}%
\special{pa 3590 1106}%
\special{pa 3595 1105}%
\special{pa 3600 1104}%
\special{sp}%
\special{pa 3600 1104}%
\special{pa 3605 1103}%
\special{pa 3610 1101}%
\special{pa 3615 1100}%
\special{pa 3620 1099}%
\special{pa 3625 1098}%
\special{pa 3630 1097}%
\special{pa 3635 1095}%
\special{pa 3640 1094}%
\special{pa 3645 1093}%
\special{pa 3650 1092}%
\special{pa 3655 1091}%
\special{pa 3660 1090}%
\special{pa 3665 1088}%
\special{pa 3670 1087}%
\special{pa 3675 1086}%
\special{pa 3680 1085}%
\special{pa 3685 1084}%
\special{pa 3690 1082}%
\special{pa 3695 1081}%
\special{pa 3700 1080}%
\special{pa 3705 1079}%
\special{pa 3710 1078}%
\special{pa 3715 1077}%
\special{pa 3720 1075}%
\special{pa 3725 1074}%
\special{pa 3730 1073}%
\special{pa 3735 1072}%
\special{pa 3740 1071}%
\special{pa 3745 1069}%
\special{pa 3750 1068}%
\special{pa 3755 1067}%
\special{pa 3760 1066}%
\special{pa 3765 1065}%
\special{pa 3770 1064}%
\special{pa 3775 1062}%
\special{pa 3780 1061}%
\special{pa 3785 1060}%
\special{pa 3790 1059}%
\special{pa 3795 1058}%
\special{pa 3800 1057}%
\special{pa 3805 1056}%
\special{pa 3810 1054}%
\special{pa 3815 1053}%
\special{pa 3820 1052}%
\special{pa 3825 1051}%
\special{pa 3830 1050}%
\special{pa 3835 1049}%
\special{pa 3840 1047}%
\special{pa 3845 1046}%
\special{pa 3850 1045}%
\special{pa 3855 1044}%
\special{pa 3860 1043}%
\special{pa 3865 1042}%
\special{pa 3870 1041}%
\special{pa 3875 1039}%
\special{pa 3880 1038}%
\special{pa 3885 1037}%
\special{pa 3890 1036}%
\special{pa 3895 1035}%
\special{pa 3900 1034}%
\special{pa 3905 1033}%
\special{pa 3910 1031}%
\special{pa 3915 1030}%
\special{pa 3920 1029}%
\special{pa 3925 1028}%
\special{pa 3930 1027}%
\special{pa 3935 1026}%
\special{pa 3940 1025}%
\special{pa 3945 1024}%
\special{pa 3950 1022}%
\special{pa 3955 1021}%
\special{pa 3960 1020}%
\special{pa 3965 1019}%
\special{pa 3970 1018}%
\special{pa 3975 1017}%
\special{pa 3980 1016}%
\special{pa 3985 1015}%
\special{pa 3990 1014}%
\special{pa 3995 1012}%
\special{pa 4000 1011}%
\special{pa 4005 1010}%
\special{pa 4010 1009}%
\special{pa 4015 1008}%
\special{pa 4020 1007}%
\special{pa 4025 1006}%
\special{pa 4030 1005}%
\special{pa 4035 1004}%
\special{pa 4040 1003}%
\special{pa 4045 1001}%
\special{pa 4050 1000}%
\special{pa 4055 999}%
\special{pa 4060 998}%
\special{pa 4065 997}%
\special{pa 4070 996}%
\special{pa 4075 995}%
\special{pa 4080 994}%
\special{pa 4085 993}%
\special{pa 4090 992}%
\special{sp}%
\special{pa 4090 992}%
\special{pa 4095 990}%
\special{pa 4100 989}%
\special{pa 4105 988}%
\special{pa 4110 987}%
\special{pa 4115 986}%
\special{pa 4120 985}%
\special{pa 4125 984}%
\special{pa 4130 983}%
\special{pa 4135 982}%
\special{pa 4140 981}%
\special{pa 4145 980}%
\special{pa 4150 979}%
\special{pa 4155 978}%
\special{pa 4160 976}%
\special{pa 4165 975}%
\special{pa 4170 974}%
\special{pa 4175 973}%
\special{pa 4180 972}%
\special{pa 4185 971}%
\special{pa 4190 970}%
\special{pa 4195 969}%
\special{pa 4200 968}%
\special{pa 4205 967}%
\special{pa 4210 966}%
\special{pa 4215 965}%
\special{pa 4220 964}%
\special{pa 4225 963}%
\special{pa 4230 962}%
\special{pa 4235 961}%
\special{pa 4240 959}%
\special{pa 4245 958}%
\special{pa 4250 957}%
\special{pa 4255 956}%
\special{pa 4260 955}%
\special{pa 4265 954}%
\special{pa 4270 953}%
\special{pa 4275 952}%
\special{pa 4280 951}%
\special{pa 4285 950}%
\special{pa 4290 949}%
\special{pa 4295 948}%
\special{pa 4300 947}%
\special{pa 4305 946}%
\special{pa 4310 945}%
\special{pa 4315 944}%
\special{pa 4320 943}%
\special{pa 4325 942}%
\special{pa 4330 941}%
\special{pa 4335 940}%
\special{pa 4340 939}%
\special{pa 4345 938}%
\special{pa 4350 937}%
\special{pa 4355 936}%
\special{pa 4360 935}%
\special{pa 4365 934}%
\special{pa 4370 933}%
\special{pa 4375 932}%
\special{pa 4380 931}%
\special{pa 4385 930}%
\special{pa 4390 929}%
\special{pa 4395 927}%
\special{pa 4400 926}%
\special{pa 4405 925}%
\special{pa 4410 924}%
\special{pa 4415 923}%
\special{pa 4420 922}%
\special{pa 4425 921}%
\special{pa 4430 920}%
\special{pa 4435 919}%
\special{pa 4440 918}%
\special{pa 4445 917}%
\special{pa 4450 916}%
\special{pa 4455 915}%
\special{pa 4460 914}%
\special{pa 4465 913}%
\special{pa 4470 912}%
\special{pa 4475 911}%
\special{pa 4480 910}%
\special{pa 4485 909}%
\special{pa 4490 908}%
\special{pa 4495 907}%
\special{pa 4500 906}%
\special{pa 4505 905}%
\special{pa 4510 904}%
\special{pa 4515 904}%
\special{pa 4520 903}%
\special{pa 4525 902}%
\special{pa 4530 901}%
\special{pa 4535 900}%
\special{pa 4540 899}%
\special{pa 4545 898}%
\special{pa 4550 897}%
\special{pa 4555 896}%
\special{pa 4560 895}%
\special{pa 4565 894}%
\special{pa 4570 893}%
\special{pa 4575 892}%
\special{pa 4580 891}%
\special{sp}%
\special{pa 4580 891}%
\special{pa 4585 890}%
\special{pa 4590 889}%
\special{pa 4595 888}%
\special{pa 4600 887}%
\special{pa 4605 886}%
\special{pa 4610 885}%
\special{pa 4615 884}%
\special{pa 4620 883}%
\special{pa 4625 882}%
\special{pa 4630 881}%
\special{pa 4635 880}%
\special{pa 4640 879}%
\special{pa 4645 878}%
\special{pa 4650 877}%
\special{pa 4655 876}%
\special{pa 4660 875}%
\special{pa 4665 874}%
\special{pa 4670 874}%
\special{pa 4675 873}%
\special{pa 4680 872}%
\special{pa 4685 871}%
\special{pa 4690 870}%
\special{pa 4695 869}%
\special{pa 4700 868}%
\special{pa 4705 867}%
\special{pa 4710 866}%
\special{pa 4715 865}%
\special{pa 4720 864}%
\special{pa 4725 863}%
\special{pa 4730 862}%
\special{pa 4735 861}%
\special{pa 4740 860}%
\special{pa 4745 859}%
\special{pa 4750 859}%
\special{pa 4755 858}%
\special{pa 4760 857}%
\special{pa 4765 856}%
\special{pa 4770 855}%
\special{pa 4775 854}%
\special{pa 4780 853}%
\special{pa 4785 852}%
\special{pa 4790 851}%
\special{pa 4795 850}%
\special{pa 4800 849}%
\special{pa 4805 848}%
\special{pa 4810 847}%
\special{pa 4815 847}%
\special{pa 4820 846}%
\special{pa 4825 845}%
\special{pa 4830 844}%
\special{pa 4835 843}%
\special{pa 4840 842}%
\special{pa 4845 841}%
\special{pa 4850 840}%
\special{pa 4855 839}%
\special{pa 4860 838}%
\special{pa 4865 837}%
\special{pa 4870 837}%
\special{pa 4875 836}%
\special{pa 4880 835}%
\special{pa 4885 834}%
\special{pa 4890 833}%
\special{pa 4895 832}%
\special{pa 4900 831}%
\special{pa 4905 830}%
\special{pa 4910 829}%
\special{pa 4915 829}%
\special{pa 4920 828}%
\special{pa 4925 827}%
\special{pa 4930 826}%
\special{pa 4935 825}%
\special{pa 4940 824}%
\special{pa 4945 823}%
\special{pa 4950 822}%
\special{pa 4955 821}%
\special{pa 4960 821}%
\special{pa 4965 820}%
\special{pa 4970 819}%
\special{pa 4975 818}%
\special{pa 4980 817}%
\special{pa 4985 816}%
\special{pa 4990 815}%
\special{pa 4995 814}%
\special{pa 5000 814}%
\special{pa 5005 813}%
\special{pa 5010 812}%
\special{pa 5015 811}%
\special{pa 5020 810}%
\special{pa 5025 809}%
\special{pa 5030 808}%
\special{pa 5035 808}%
\special{pa 5040 807}%
\special{pa 5045 806}%
\special{pa 5050 805}%
\special{pa 5055 804}%
\special{pa 5060 803}%
\special{pa 5065 802}%
\special{pa 5070 802}%
\special{sp}%
\special{pa 5070 802}%
\special{pa 5075 801}%
\special{pa 5080 800}%
\special{pa 5085 799}%
\special{pa 5090 798}%
\special{pa 5095 797}%
\special{pa 5100 796}%
\special{pa 5105 796}%
\special{pa 5110 795}%
\special{pa 5115 794}%
\special{pa 5120 793}%
\special{pa 5125 792}%
\special{pa 5130 791}%
\special{pa 5135 791}%
\special{pa 5140 790}%
\special{pa 5145 789}%
\special{pa 5150 788}%
\special{pa 5155 787}%
\special{pa 5160 786}%
\special{pa 5165 786}%
\special{pa 5170 785}%
\special{pa 5175 784}%
\special{pa 5180 783}%
\special{pa 5185 782}%
\special{pa 5190 781}%
\special{pa 5195 781}%
\special{pa 5200 780}%
\special{pa 5205 779}%
\special{pa 5210 778}%
\special{pa 5215 777}%
\special{pa 5220 777}%
\special{pa 5225 776}%
\special{pa 5230 775}%
\special{pa 5235 774}%
\special{pa 5240 773}%
\special{pa 5245 772}%
\special{pa 5250 772}%
\special{pa 5255 771}%
\special{pa 5260 770}%
\special{pa 5265 769}%
\special{pa 5270 768}%
\special{pa 5275 768}%
\special{sp}%
% FUNC 2 0 3 0
% 9 1644 920 5277 3090 1644 3090 5056 3090 1644 1377 1644 920 5275 3090 0 3 0 0
% (0.0326603-0.033+0.000603233*(12*x+1)-0.00000598045*(12*x+1)^2)/0.006
\special{pn 8}%
\special{pa 1640 2619}%
\special{pa 1645 2616}%
\special{pa 1650 2613}%
\special{pa 1655 2610}%
\special{pa 1660 2607}%
\special{pa 1665 2604}%
\special{pa 1670 2601}%
\special{pa 1675 2598}%
\special{pa 1680 2595}%
\special{pa 1685 2592}%
\special{pa 1690 2589}%
\special{pa 1695 2586}%
\special{pa 1700 2583}%
\special{pa 1705 2580}%
\special{pa 1710 2577}%
\special{pa 1715 2574}%
\special{pa 1720 2571}%
\special{pa 1725 2569}%
\special{pa 1730 2566}%
\special{pa 1735 2563}%
\special{pa 1740 2560}%
\special{pa 1745 2557}%
\special{pa 1750 2554}%
\special{pa 1755 2551}%
\special{pa 1760 2548}%
\special{pa 1765 2545}%
\special{pa 1770 2542}%
\special{pa 1775 2539}%
\special{pa 1780 2536}%
\special{pa 1785 2533}%
\special{pa 1790 2530}%
\special{pa 1795 2527}%
\special{pa 1800 2524}%
\special{pa 1805 2521}%
\special{pa 1810 2518}%
\special{pa 1815 2516}%
\special{pa 1820 2513}%
\special{pa 1825 2510}%
\special{pa 1830 2507}%
\special{pa 1835 2504}%
\special{pa 1840 2501}%
\special{pa 1845 2498}%
\special{pa 1850 2495}%
\special{pa 1855 2492}%
\special{pa 1860 2489}%
\special{pa 1865 2486}%
\special{pa 1870 2483}%
\special{pa 1875 2480}%
\special{pa 1880 2478}%
\special{pa 1885 2475}%
\special{pa 1890 2472}%
\special{pa 1895 2469}%
\special{pa 1900 2466}%
\special{pa 1905 2463}%
\special{pa 1910 2460}%
\special{pa 1915 2457}%
\special{pa 1920 2454}%
\special{pa 1925 2451}%
\special{pa 1930 2448}%
\special{pa 1935 2445}%
\special{pa 1940 2443}%
\special{pa 1945 2440}%
\special{pa 1950 2437}%
\special{pa 1955 2434}%
\special{pa 1960 2431}%
\special{pa 1965 2428}%
\special{pa 1970 2425}%
\special{pa 1975 2422}%
\special{pa 1980 2419}%
\special{pa 1985 2416}%
\special{pa 1990 2414}%
\special{pa 1995 2411}%
\special{pa 2000 2408}%
\special{pa 2005 2405}%
\special{pa 2010 2402}%
\special{pa 2015 2399}%
\special{pa 2020 2396}%
\special{pa 2025 2393}%
\special{pa 2030 2390}%
\special{pa 2035 2388}%
\special{pa 2040 2385}%
\special{pa 2045 2382}%
\special{pa 2050 2379}%
\special{pa 2055 2376}%
\special{pa 2060 2373}%
\special{pa 2065 2370}%
\special{pa 2070 2367}%
\special{pa 2075 2365}%
\special{pa 2080 2362}%
\special{pa 2085 2359}%
\special{pa 2090 2356}%
\special{pa 2095 2353}%
\special{pa 2100 2350}%
\special{pa 2105 2347}%
\special{pa 2110 2344}%
\special{pa 2115 2342}%
\special{pa 2120 2339}%
\special{pa 2125 2336}%
\special{pa 2130 2333}%
\special{sp}%
\special{pa 2130 2333}%
\special{pa 2135 2330}%
\special{pa 2140 2327}%
\special{pa 2145 2324}%
\special{pa 2150 2321}%
\special{pa 2155 2319}%
\special{pa 2160 2316}%
\special{pa 2165 2313}%
\special{pa 2170 2310}%
\special{pa 2175 2307}%
\special{pa 2180 2304}%
\special{pa 2185 2301}%
\special{pa 2190 2299}%
\special{pa 2195 2296}%
\special{pa 2200 2293}%
\special{pa 2205 2290}%
\special{pa 2210 2287}%
\special{pa 2215 2284}%
\special{pa 2220 2282}%
\special{pa 2225 2279}%
\special{pa 2230 2276}%
\special{pa 2235 2273}%
\special{pa 2240 2270}%
\special{pa 2245 2267}%
\special{pa 2250 2264}%
\special{pa 2255 2262}%
\special{pa 2260 2259}%
\special{pa 2265 2256}%
\special{pa 2270 2253}%
\special{pa 2275 2250}%
\special{pa 2280 2247}%
\special{pa 2285 2245}%
\special{pa 2290 2242}%
\special{pa 2295 2239}%
\special{pa 2300 2236}%
\special{pa 2305 2233}%
\special{pa 2310 2230}%
\special{pa 2315 2228}%
\special{pa 2320 2225}%
\special{pa 2325 2222}%
\special{pa 2330 2219}%
\special{pa 2335 2216}%
\special{pa 2340 2213}%
\special{pa 2345 2211}%
\special{pa 2350 2208}%
\special{pa 2355 2205}%
\special{pa 2360 2202}%
\special{pa 2365 2199}%
\special{pa 2370 2197}%
\special{pa 2375 2194}%
\special{pa 2380 2191}%
\special{pa 2385 2188}%
\special{pa 2390 2185}%
\special{pa 2395 2183}%
\special{pa 2400 2180}%
\special{pa 2405 2177}%
\special{pa 2410 2174}%
\special{pa 2415 2171}%
\special{pa 2420 2168}%
\special{pa 2425 2166}%
\special{pa 2430 2163}%
\special{pa 2435 2160}%
\special{pa 2440 2157}%
\special{pa 2445 2154}%
\special{pa 2450 2152}%
\special{pa 2455 2149}%
\special{pa 2460 2146}%
\special{pa 2465 2143}%
\special{pa 2470 2140}%
\special{pa 2475 2138}%
\special{pa 2480 2135}%
\special{pa 2485 2132}%
\special{pa 2490 2129}%
\special{pa 2495 2127}%
\special{pa 2500 2124}%
\special{pa 2505 2121}%
\special{pa 2510 2118}%
\special{pa 2515 2115}%
\special{pa 2520 2113}%
\special{pa 2525 2110}%
\special{pa 2530 2107}%
\special{pa 2535 2104}%
\special{pa 2540 2101}%
\special{pa 2545 2099}%
\special{pa 2550 2096}%
\special{pa 2555 2093}%
\special{pa 2560 2090}%
\special{pa 2565 2088}%
\special{pa 2570 2085}%
\special{pa 2575 2082}%
\special{pa 2580 2079}%
\special{pa 2585 2076}%
\special{pa 2590 2074}%
\special{pa 2595 2071}%
\special{pa 2600 2068}%
\special{pa 2605 2065}%
\special{pa 2610 2063}%
\special{pa 2615 2060}%
\special{pa 2620 2057}%
\special{sp}%
\special{pa 2620 2057}%
\special{pa 2625 2054}%
\special{pa 2630 2052}%
\special{pa 2635 2049}%
\special{pa 2640 2046}%
\special{pa 2645 2043}%
\special{pa 2650 2041}%
\special{pa 2655 2038}%
\special{pa 2660 2035}%
\special{pa 2665 2032}%
\special{pa 2670 2030}%
\special{pa 2675 2027}%
\special{pa 2680 2024}%
\special{pa 2685 2021}%
\special{pa 2690 2019}%
\special{pa 2695 2016}%
\special{pa 2700 2013}%
\special{pa 2705 2010}%
\special{pa 2710 2008}%
\special{pa 2715 2005}%
\special{pa 2720 2002}%
\special{pa 2725 1999}%
\special{pa 2730 1997}%
\special{pa 2735 1994}%
\special{pa 2740 1991}%
\special{pa 2745 1988}%
\special{pa 2750 1986}%
\special{pa 2755 1983}%
\special{pa 2760 1980}%
\special{pa 2765 1977}%
\special{pa 2770 1975}%
\special{pa 2775 1972}%
\special{pa 2780 1969}%
\special{pa 2785 1967}%
\special{pa 2790 1964}%
\special{pa 2795 1961}%
\special{pa 2800 1958}%
\special{pa 2805 1956}%
\special{pa 2810 1953}%
\special{pa 2815 1950}%
\special{pa 2820 1947}%
\special{pa 2825 1945}%
\special{pa 2830 1942}%
\special{pa 2835 1939}%
\special{pa 2840 1937}%
\special{pa 2845 1934}%
\special{pa 2850 1931}%
\special{pa 2855 1928}%
\special{pa 2860 1926}%
\special{pa 2865 1923}%
\special{pa 2870 1920}%
\special{pa 2875 1918}%
\special{pa 2880 1915}%
\special{pa 2885 1912}%
\special{pa 2890 1910}%
\special{pa 2895 1907}%
\special{pa 2900 1904}%
\special{pa 2905 1901}%
\special{pa 2910 1899}%
\special{pa 2915 1896}%
\special{pa 2920 1893}%
\special{pa 2925 1891}%
\special{pa 2930 1888}%
\special{pa 2935 1885}%
\special{pa 2940 1883}%
\special{pa 2945 1880}%
\special{pa 2950 1877}%
\special{pa 2955 1874}%
\special{pa 2960 1872}%
\special{pa 2965 1869}%
\special{pa 2970 1866}%
\special{pa 2975 1864}%
\special{pa 2980 1861}%
\special{pa 2985 1858}%
\special{pa 2990 1856}%
\special{pa 2995 1853}%
\special{pa 3000 1850}%
\special{pa 3005 1848}%
\special{pa 3010 1845}%
\special{pa 3015 1842}%
\special{pa 3020 1840}%
\special{pa 3025 1837}%
\special{pa 3030 1834}%
\special{pa 3035 1831}%
\special{pa 3040 1829}%
\special{pa 3045 1826}%
\special{pa 3050 1823}%
\special{pa 3055 1821}%
\special{pa 3060 1818}%
\special{pa 3065 1815}%
\special{pa 3070 1813}%
\special{pa 3075 1810}%
\special{pa 3080 1807}%
\special{pa 3085 1805}%
\special{pa 3090 1802}%
\special{pa 3095 1799}%
\special{pa 3100 1797}%
\special{pa 3105 1794}%
\special{pa 3110 1791}%
\special{sp}%
\special{pa 3110 1791}%
\special{pa 3115 1789}%
\special{pa 3120 1786}%
\special{pa 3125 1784}%
\special{pa 3130 1781}%
\special{pa 3135 1778}%
\special{pa 3140 1776}%
\special{pa 3145 1773}%
\special{pa 3150 1770}%
\special{pa 3155 1768}%
\special{pa 3160 1765}%
\special{pa 3165 1762}%
\special{pa 3170 1760}%
\special{pa 3175 1757}%
\special{pa 3180 1754}%
\special{pa 3185 1752}%
\special{pa 3190 1749}%
\special{pa 3195 1746}%
\special{pa 3200 1744}%
\special{pa 3205 1741}%
\special{pa 3210 1739}%
\special{pa 3215 1736}%
\special{pa 3220 1733}%
\special{pa 3225 1731}%
\special{pa 3230 1728}%
\special{pa 3235 1725}%
\special{pa 3240 1723}%
\special{pa 3245 1720}%
\special{pa 3250 1717}%
\special{pa 3255 1715}%
\special{pa 3260 1712}%
\special{pa 3265 1710}%
\special{pa 3270 1707}%
\special{pa 3275 1704}%
\special{pa 3280 1702}%
\special{pa 3285 1699}%
\special{pa 3290 1696}%
\special{pa 3295 1694}%
\special{pa 3300 1691}%
\special{pa 3305 1689}%
\special{pa 3310 1686}%
\special{pa 3315 1683}%
\special{pa 3320 1681}%
\special{pa 3325 1678}%
\special{pa 3330 1676}%
\special{pa 3335 1673}%
\special{pa 3340 1670}%
\special{pa 3345 1668}%
\special{pa 3350 1665}%
\special{pa 3355 1662}%
\special{pa 3360 1660}%
\special{pa 3365 1657}%
\special{pa 3370 1655}%
\special{pa 3375 1652}%
\special{pa 3380 1649}%
\special{pa 3385 1647}%
\special{pa 3390 1644}%
\special{pa 3395 1642}%
\special{pa 3400 1639}%
\special{pa 3405 1636}%
\special{pa 3410 1634}%
\special{pa 3415 1631}%
\special{pa 3420 1629}%
\special{pa 3425 1626}%
\special{pa 3430 1623}%
\special{pa 3435 1621}%
\special{pa 3440 1618}%
\special{pa 3445 1616}%
\special{pa 3450 1613}%
\special{pa 3455 1611}%
\special{pa 3460 1608}%
\special{pa 3465 1605}%
\special{pa 3470 1603}%
\special{pa 3475 1600}%
\special{pa 3480 1598}%
\special{pa 3485 1595}%
\special{pa 3490 1592}%
\special{pa 3495 1590}%
\special{pa 3500 1587}%
\special{pa 3505 1585}%
\special{pa 3510 1582}%
\special{pa 3515 1580}%
\special{pa 3520 1577}%
\special{pa 3525 1574}%
\special{pa 3530 1572}%
\special{pa 3535 1569}%
\special{pa 3540 1567}%
\special{pa 3545 1564}%
\special{pa 3550 1562}%
\special{pa 3555 1559}%
\special{pa 3560 1556}%
\special{pa 3565 1554}%
\special{pa 3570 1551}%
\special{pa 3575 1549}%
\special{pa 3580 1546}%
\special{pa 3585 1544}%
\special{pa 3590 1541}%
\special{pa 3595 1539}%
\special{pa 3600 1536}%
\special{sp}%
\special{pa 3600 1536}%
\special{pa 3605 1533}%
\special{pa 3610 1531}%
\special{pa 3615 1528}%
\special{pa 3620 1526}%
\special{pa 3625 1523}%
\special{pa 3630 1521}%
\special{pa 3635 1518}%
\special{pa 3640 1516}%
\special{pa 3645 1513}%
\special{pa 3650 1510}%
\special{pa 3655 1508}%
\special{pa 3660 1505}%
\special{pa 3665 1503}%
\special{pa 3670 1500}%
\special{pa 3675 1498}%
\special{pa 3680 1495}%
\special{pa 3685 1493}%
\special{pa 3690 1490}%
\special{pa 3695 1488}%
\special{pa 3700 1485}%
\special{pa 3705 1483}%
\special{pa 3710 1480}%
\special{pa 3715 1478}%
\special{pa 3720 1475}%
\special{pa 3725 1472}%
\special{pa 3730 1470}%
\special{pa 3735 1467}%
\special{pa 3740 1465}%
\special{pa 3745 1462}%
\special{pa 3750 1460}%
\special{pa 3755 1457}%
\special{pa 3760 1455}%
\special{pa 3765 1452}%
\special{pa 3770 1450}%
\special{pa 3775 1447}%
\special{pa 3780 1445}%
\special{pa 3785 1442}%
\special{pa 3790 1440}%
\special{pa 3795 1437}%
\special{pa 3800 1435}%
\special{pa 3805 1432}%
\special{pa 3810 1430}%
\special{pa 3815 1427}%
\special{pa 3820 1425}%
\special{pa 3825 1422}%
\special{pa 3830 1420}%
\special{pa 3835 1417}%
\special{pa 3840 1415}%
\special{pa 3845 1412}%
\special{pa 3850 1410}%
\special{pa 3855 1407}%
\special{pa 3860 1405}%
\special{pa 3865 1402}%
\special{pa 3870 1400}%
\special{pa 3875 1397}%
\special{pa 3880 1395}%
\special{pa 3885 1392}%
\special{pa 3890 1390}%
\special{pa 3895 1387}%
\special{pa 3900 1385}%
\special{pa 3905 1382}%
\special{pa 3910 1380}%
\special{pa 3915 1377}%
\special{pa 3920 1375}%
\special{pa 3925 1372}%
\special{pa 3930 1370}%
\special{pa 3935 1367}%
\special{pa 3940 1365}%
\special{pa 3945 1362}%
\special{pa 3950 1360}%
\special{pa 3955 1357}%
\special{pa 3960 1355}%
\special{pa 3965 1352}%
\special{pa 3970 1350}%
\special{pa 3975 1347}%
\special{pa 3980 1345}%
\special{pa 3985 1342}%
\special{pa 3990 1340}%
\special{pa 3995 1337}%
\special{pa 4000 1335}%
\special{pa 4005 1332}%
\special{pa 4010 1330}%
\special{pa 4015 1328}%
\special{pa 4020 1325}%
\special{pa 4025 1323}%
\special{pa 4030 1320}%
\special{pa 4035 1318}%
\special{pa 4040 1315}%
\special{pa 4045 1313}%
\special{pa 4050 1310}%
\special{pa 4055 1308}%
\special{pa 4060 1305}%
\special{pa 4065 1303}%
\special{pa 4070 1300}%
\special{pa 4075 1298}%
\special{pa 4080 1296}%
\special{pa 4085 1293}%
\special{pa 4090 1291}%
\special{sp}%
\special{pa 4090 1291}%
\special{pa 4095 1288}%
\special{pa 4100 1286}%
\special{pa 4105 1283}%
\special{pa 4110 1281}%
\special{pa 4115 1278}%
\special{pa 4120 1276}%
\special{pa 4125 1274}%
\special{pa 4130 1271}%
\special{pa 4135 1269}%
\special{pa 4140 1266}%
\special{pa 4145 1264}%
\special{pa 4150 1261}%
\special{pa 4155 1259}%
\special{pa 4160 1256}%
\special{pa 4165 1254}%
\special{pa 4170 1252}%
\special{pa 4175 1249}%
\special{pa 4180 1247}%
\special{pa 4185 1244}%
\special{pa 4190 1242}%
\special{pa 4195 1239}%
\special{pa 4200 1237}%
\special{pa 4205 1235}%
\special{pa 4210 1232}%
\special{pa 4215 1230}%
\special{pa 4220 1227}%
\special{pa 4225 1225}%
\special{pa 4230 1222}%
\special{pa 4235 1220}%
\special{pa 4240 1218}%
\special{pa 4245 1215}%
\special{pa 4250 1213}%
\special{pa 4255 1210}%
\special{pa 4260 1208}%
\special{pa 4265 1205}%
\special{pa 4270 1203}%
\special{pa 4275 1201}%
\special{pa 4280 1198}%
\special{pa 4285 1196}%
\special{pa 4290 1193}%
\special{pa 4295 1191}%
\special{pa 4300 1189}%
\special{pa 4305 1186}%
\special{pa 4310 1184}%
\special{pa 4315 1181}%
\special{pa 4320 1179}%
\special{pa 4325 1177}%
\special{pa 4330 1174}%
\special{pa 4335 1172}%
\special{pa 4340 1169}%
\special{pa 4345 1167}%
\special{pa 4350 1165}%
\special{pa 4355 1162}%
\special{pa 4360 1160}%
\special{pa 4365 1157}%
\special{pa 4370 1155}%
\special{pa 4375 1153}%
\special{pa 4380 1150}%
\special{pa 4385 1148}%
\special{pa 4390 1145}%
\special{pa 4395 1143}%
\special{pa 4400 1141}%
\special{pa 4405 1138}%
\special{pa 4410 1136}%
\special{pa 4415 1133}%
\special{pa 4420 1131}%
\special{pa 4425 1129}%
\special{pa 4430 1126}%
\special{pa 4435 1124}%
\special{pa 4440 1122}%
\special{pa 4445 1119}%
\special{pa 4450 1117}%
\special{pa 4455 1114}%
\special{pa 4460 1112}%
\special{pa 4465 1110}%
\special{pa 4470 1107}%
\special{pa 4475 1105}%
\special{pa 4480 1103}%
\special{pa 4485 1100}%
\special{pa 4490 1098}%
\special{pa 4495 1095}%
\special{pa 4500 1093}%
\special{pa 4505 1091}%
\special{pa 4510 1088}%
\special{pa 4515 1086}%
\special{pa 4520 1084}%
\special{pa 4525 1081}%
\special{pa 4530 1079}%
\special{pa 4535 1077}%
\special{pa 4540 1074}%
\special{pa 4545 1072}%
\special{pa 4550 1070}%
\special{pa 4555 1067}%
\special{pa 4560 1065}%
\special{pa 4565 1062}%
\special{pa 4570 1060}%
\special{pa 4575 1058}%
\special{pa 4580 1055}%
\special{sp}%
\special{pa 4580 1055}%
\special{pa 4585 1053}%
\special{pa 4590 1051}%
\special{pa 4595 1048}%
\special{pa 4600 1046}%
\special{pa 4605 1044}%
\special{pa 4610 1041}%
\special{pa 4615 1039}%
\special{pa 4620 1037}%
\special{pa 4625 1034}%
\special{pa 4630 1032}%
\special{pa 4635 1030}%
\special{pa 4640 1027}%
\special{pa 4645 1025}%
\special{pa 4650 1023}%
\special{pa 4655 1020}%
\special{pa 4660 1018}%
\special{pa 4665 1016}%
\special{pa 4670 1013}%
\special{pa 4675 1011}%
\special{pa 4680 1009}%
\special{pa 4685 1006}%
\special{pa 4690 1004}%
\special{pa 4695 1002}%
\special{pa 4700 999}%
\special{pa 4705 997}%
\special{pa 4710 995}%
\special{pa 4715 992}%
\special{pa 4720 990}%
\special{pa 4725 988}%
\special{pa 4730 985}%
\special{pa 4735 983}%
\special{pa 4740 981}%
\special{pa 4745 979}%
\special{pa 4750 976}%
\special{pa 4755 974}%
\special{pa 4760 972}%
\special{pa 4765 969}%
\special{pa 4770 967}%
\special{pa 4775 965}%
\special{pa 4780 962}%
\special{pa 4785 960}%
\special{pa 4790 958}%
\special{pa 4795 955}%
\special{pa 4800 953}%
\special{pa 4805 951}%
\special{pa 4810 949}%
\special{pa 4815 946}%
\special{pa 4820 944}%
\special{pa 4825 942}%
\special{pa 4830 939}%
\special{pa 4835 937}%
\special{pa 4840 935}%
\special{pa 4845 932}%
\special{pa 4850 930}%
\special{pa 4855 928}%
\special{pa 4860 926}%
\special{pa 4865 923}%
\special{pa 4870 921}%
\special{pa 4875 919}%
\special{pa 4880 916}%
\special{pa 4885 914}%
\special{pa 4890 912}%
\special{pa 4895 910}%
\special{pa 4900 907}%
\special{pa 4905 905}%
\special{pa 4910 903}%
\special{pa 4915 900}%
\special{pa 4920 898}%
\special{pa 4925 896}%
\special{pa 4930 894}%
\special{pa 4935 891}%
\special{pa 4940 889}%
\special{pa 4945 887}%
\special{pa 4950 885}%
\special{pa 4955 882}%
\special{pa 4960 880}%
\special{pa 4965 878}%
\special{pa 4970 875}%
\special{pa 4975 873}%
\special{pa 4980 871}%
\special{pa 4985 869}%
\special{pa 4990 866}%
\special{pa 4995 864}%
\special{pa 5000 862}%
\special{pa 5005 860}%
\special{pa 5010 857}%
\special{pa 5015 855}%
\special{pa 5020 853}%
\special{pa 5025 851}%
\special{pa 5030 848}%
\special{pa 5035 846}%
\special{pa 5040 844}%
\special{pa 5045 842}%
\special{pa 5050 839}%
\special{pa 5055 837}%
\special{pa 5060 835}%
\special{pa 5065 833}%
\special{pa 5070 830}%
\special{sp}%
\special{pa 5070 830}%
\special{pa 5075 828}%
\special{pa 5080 826}%
\special{pa 5085 824}%
\special{pa 5090 821}%
\special{pa 5095 819}%
\special{pa 5100 817}%
\special{pa 5105 815}%
\special{pa 5110 812}%
\special{pa 5115 810}%
\special{pa 5120 808}%
\special{pa 5125 806}%
\special{pa 5130 803}%
\special{pa 5135 801}%
\special{pa 5140 799}%
\special{pa 5145 797}%
\special{pa 5150 795}%
\special{pa 5155 792}%
\special{pa 5160 790}%
\special{pa 5165 788}%
\special{pa 5170 786}%
\special{pa 5175 783}%
\special{pa 5180 781}%
\special{pa 5185 779}%
\special{pa 5190 777}%
\special{pa 5195 775}%
\special{pa 5200 772}%
\special{pa 5205 770}%
\special{pa 5210 768}%
\special{pa 5215 766}%
\special{pa 5220 763}%
\special{pa 5225 761}%
\special{pa 5230 759}%
\special{pa 5235 757}%
\special{pa 5240 755}%
\special{pa 5245 752}%
\special{pa 5250 750}%
\special{pa 5255 748}%
\special{pa 5260 746}%
\special{pa 5265 744}%
\special{pa 5270 741}%
\special{pa 5275 739}%
\special{sp}%
% STR 2 0 3 0
% 3 640 1958 640 2030 33 206
% $\Delta r$
\put(6.4000,-16.3000){\makebox(0,0)[lb]{$\Delta r$}}%
% STR 2 0 3 0
% 3 2440 3779 2440 3850 33 206
% $m_{\phi}$
\put(24.4000,-34.5000){\makebox(0,0)[lb]{$m_{\phi}$}}%
% STR 2 0 3 0
% 3 3260 3789 3260 3860 33 206
% $m_{h}=m_{H}$
\put(32.6000,-34.6000){\makebox(0,0)[lb]{$m_{h}=m_{H}$}}%
% STR 2 0 3 0
% 3 4630 3799 4630 3870 33 206
% GeV
\put(46.3000,-34.7000){\makebox(0,0)[lb]{GeV}}%
% STR 2 0 3 0
% 3 4550 1560 4550 1660 2 0
% THDM
\put(45.5000,-12.6000){\makebox(0,0)[lb]{THDM}}%
% STR 2 0 3 0
% 3 2690 1490 2690 1590 2 0
% SM
\put(26.9000,-11.9000){\makebox(0,0)[lb]{SM}}%
% POLYGON 2 0 3 0
% 5 1640 3370 5390 3370 5390 960 1640 960 1640 3370
% 
\special{pn 8}%
\special{pa 1640 2970}%
\special{pa 5390 2970}%
\special{pa 5390 560}%
\special{pa 1640 560}%
\special{pa 1640 2970}%
\special{fp}%
% LINE 2 0 3 0
% 2 1930 3370 1930 3313
% 
\special{pn 8}%
\special{pa 1930 2970}%
\special{pa 1930 2913}%
\special{fp}%
% LINE 2 0 3 0
% 2 2500 3370 2500 3313
% 
\special{pn 8}%
\special{pa 2500 2970}%
\special{pa 2500 2913}%
\special{fp}%
% LINE 2 0 3 0
% 2 3070 3370 3070 3313
% 
\special{pn 8}%
\special{pa 3070 2970}%
\special{pa 3070 2913}%
\special{fp}%
% LINE 2 0 3 0
% 2 3630 3370 3630 3313
% 
\special{pn 8}%
\special{pa 3630 2970}%
\special{pa 3630 2913}%
\special{fp}%
% LINE 2 0 3 0
% 2 4200 3370 4200 3313
% 
\special{pn 8}%
\special{pa 4200 2970}%
\special{pa 4200 2913}%
\special{fp}%
% LINE 2 0 3 0
% 2 4770 3370 4770 3313
% 
\special{pn 8}%
\special{pa 4770 2970}%
\special{pa 4770 2913}%
\special{fp}%
% LINE 2 0 3 0
% 2 1930 960 1930 1017
% 
\special{pn 8}%
\special{pa 1930 560}%
\special{pa 1930 617}%
\special{fp}%
% LINE 2 0 3 0
% 2 2500 960 2500 1017
% 
\special{pn 8}%
\special{pa 2500 560}%
\special{pa 2500 617}%
\special{fp}%
% LINE 2 0 3 0
% 2 3070 960 3070 1017
% 
\special{pn 8}%
\special{pa 3070 560}%
\special{pa 3070 617}%
\special{fp}%
% LINE 2 0 3 0
% 2 3630 960 3630 1017
% 
\special{pn 8}%
\special{pa 3630 560}%
\special{pa 3630 617}%
\special{fp}%
% LINE 2 0 3 0
% 2 4200 960 4200 1017
% 
\special{pn 8}%
\special{pa 4200 560}%
\special{pa 4200 617}%
\special{fp}%
% LINE 2 0 3 0
% 2 4770 960 4770 1017
% 
\special{pn 8}%
\special{pa 4770 560}%
\special{pa 4770 617}%
\special{fp}%
% LINE 2 0 3 0
% 2 2210 3370 2210 3270
% 
\special{pn 8}%
\special{pa 2210 2970}%
\special{pa 2210 2870}%
\special{fp}%
% LINE 2 0 3 0
% 2 2780 3370 2780 3270
% 
\special{pn 8}%
\special{pa 2780 2970}%
\special{pa 2780 2870}%
\special{fp}%
% LINE 2 0 3 0
% 2 3350 3370 3350 3270
% 
\special{pn 8}%
\special{pa 3350 2970}%
\special{pa 3350 2870}%
\special{fp}%
% LINE 2 0 3 0
% 2 3920 3370 3920 3270
% 
\special{pn 8}%
\special{pa 3920 2970}%
\special{pa 3920 2870}%
\special{fp}%
% LINE 2 0 3 0
% 2 4490 3370 4490 3270
% 
\special{pn 8}%
\special{pa 4490 2970}%
\special{pa 4490 2870}%
\special{fp}%
% LINE 2 0 3 0
% 2 5050 3370 5050 3270
% 
\special{pn 8}%
\special{pa 5050 2970}%
\special{pa 5050 2870}%
\special{fp}%
% LINE 2 0 3 0
% 2 2210 960 2210 1060
% 
\special{pn 8}%
\special{pa 2210 560}%
\special{pa 2210 660}%
\special{fp}%
% LINE 2 0 3 0
% 2 2780 960 2780 1060
% 
\special{pn 8}%
\special{pa 2780 560}%
\special{pa 2780 660}%
\special{fp}%
% LINE 2 0 3 0
% 2 3350 960 3350 1060
% 
\special{pn 8}%
\special{pa 3350 560}%
\special{pa 3350 660}%
\special{fp}%
% LINE 2 0 3 0
% 2 3920 960 3920 1060
% 
\special{pn 8}%
\special{pa 3920 560}%
\special{pa 3920 660}%
\special{fp}%
% LINE 2 0 3 0
% 2 4490 960 4490 1060
% 
\special{pn 8}%
\special{pa 4490 560}%
\special{pa 4490 660}%
\special{fp}%
% LINE 2 0 3 0
% 2 5050 960 5050 1060
% 
\special{pn 8}%
\special{pa 5050 560}%
\special{pa 5050 660}%
\special{fp}%
% LINE 2 0 3 0
% 2 5390 1380 5326 1380
% 
\special{pn 8}%
\special{pa 5390 980}%
\special{pa 5326 980}%
\special{fp}%
% LINE 2 0 3 0
% 2 5390 1660 5326 1660
% 
\special{pn 8}%
\special{pa 5390 1260}%
\special{pa 5326 1260}%
\special{fp}%
% LINE 2 0 3 0
% 2 5390 1950 5326 1950
% 
\special{pn 8}%
\special{pa 5390 1550}%
\special{pa 5326 1550}%
\special{fp}%
% LINE 2 0 3 0
% 2 5390 2220 5326 2220
% 
\special{pn 8}%
\special{pa 5390 1820}%
\special{pa 5326 1820}%
\special{fp}%
% LINE 2 0 3 0
% 2 5390 2520 5326 2520
% 
\special{pn 8}%
\special{pa 5390 2120}%
\special{pa 5326 2120}%
\special{fp}%
% LINE 2 0 3 0
% 2 5390 2800 5326 2800
% 
\special{pn 8}%
\special{pa 5390 2400}%
\special{pa 5326 2400}%
\special{fp}%
% LINE 2 0 3 0
% 2 5390 3090 5326 3090
% 
\special{pn 8}%
\special{pa 5390 2690}%
\special{pa 5326 2690}%
\special{fp}%
% STR 2 0 3 0
% 3 4450 1780 4450 1880 2 0
% (example)
\put(44.5000,-14.8000){\makebox(0,0)[lb]{(example)}}%
% LINE 3 0 3 0
% 60 2670 3000 2630 2920 2715 3000 2630 2830 2760 3000 2630 2740 2805 3000 2630 2650 2850 3000 2630 2560 2895 3000 2630 2470 2940 3000 2630 2380 2985 3000 2630 2290 3030 3000 2630 2200 3075 3000 2630 2110 3120 3000 2630 2020 3165 3000 2640 1950 3210 3000 2685 1950 3255 3000 2730 1950 3300 3000 2775 1950 3345 3000 2820 1950 3390 3000 2865 1950 3435 3000 2910 1950 3480 3000 2955 1950 3525 3000 3000 1950 3570 3000 3045 1950 3615 3000 3090 1950 3660 3000 3135 1950 3705 3000 3180 1950 3750 3000 3225 1950 3795 3000 3270 1950 3840 3000 3315 1950 3885 3000 3360 1950 3930 3000 3405 1950 3975 3000 3450 1950
% 
\special{pn 4}%
\special{pa 2670 2600}%
\special{pa 2630 2520}%
\special{fp}%
\special{pa 2715 2600}%
\special{pa 2630 2430}%
\special{fp}%
\special{pa 2760 2600}%
\special{pa 2630 2340}%
\special{fp}%
\special{pa 2805 2600}%
\special{pa 2630 2250}%
\special{fp}%
\special{pa 2850 2600}%
\special{pa 2630 2160}%
\special{fp}%
\special{pa 2895 2600}%
\special{pa 2630 2070}%
\special{fp}%
\special{pa 2940 2600}%
\special{pa 2630 1980}%
\special{fp}%
\special{pa 2985 2600}%
\special{pa 2630 1890}%
\special{fp}%
\special{pa 3030 2600}%
\special{pa 2630 1800}%
\special{fp}%
\special{pa 3075 2600}%
\special{pa 2630 1710}%
\special{fp}%
\special{pa 3120 2600}%
\special{pa 2630 1620}%
\special{fp}%
\special{pa 3165 2600}%
\special{pa 2640 1550}%
\special{fp}%
\special{pa 3210 2600}%
\special{pa 2685 1550}%
\special{fp}%
\special{pa 3255 2600}%
\special{pa 2730 1550}%
\special{fp}%
\special{pa 3300 2600}%
\special{pa 2775 1550}%
\special{fp}%
\special{pa 3345 2600}%
\special{pa 2820 1550}%
\special{fp}%
\special{pa 3390 2600}%
\special{pa 2865 1550}%
\special{fp}%
\special{pa 3435 2600}%
\special{pa 2910 1550}%
\special{fp}%
\special{pa 3480 2600}%
\special{pa 2955 1550}%
\special{fp}%
\special{pa 3525 2600}%
\special{pa 3000 1550}%
\special{fp}%
\special{pa 3570 2600}%
\special{pa 3045 1550}%
\special{fp}%
\special{pa 3615 2600}%
\special{pa 3090 1550}%
\special{fp}%
\special{pa 3660 2600}%
\special{pa 3135 1550}%
\special{fp}%
\special{pa 3705 2600}%
\special{pa 3180 1550}%
\special{fp}%
\special{pa 3750 2600}%
\special{pa 3225 1550}%
\special{fp}%
\special{pa 3795 2600}%
\special{pa 3270 1550}%
\special{fp}%
\special{pa 3840 2600}%
\special{pa 3315 1550}%
\special{fp}%
\special{pa 3885 2600}%
\special{pa 3360 1550}%
\special{fp}%
\special{pa 3930 2600}%
\special{pa 3405 1550}%
\special{fp}%
\special{pa 3975 2600}%
\special{pa 3450 1550}%
\special{fp}%
% LINE 3 0 3 1
% 60 4020 3000 3495 1950 4065 3000 3540 1950 4110 3000 3585 1950 4155 3000 3630 1950 4200 3000 3675 1950 4245 3000 3720 1950 4290 3000 3765 1950 4335 3000 3810 1950 4380 3000 3855 1950 4425 3000 3900 1950 4470 3000 3945 1950 4515 3000 3990 1950 4560 3000 4035 1950 4605 3000 4080 1950 4650 3000 4125 1950 4695 3000 4170 1950 4740 3000 4215 1950 4785 3000 4260 1950 4830 3000 4305 1950 4875 3000 4350 1950 4920 3000 4395 1950 4965 3000 4440 1950 5010 3000 4485 1950 5010 2910 4530 1950 5010 2820 4575 1950 5010 2730 4620 1950 5010 2640 4665 1950 5010 2550 4710 1950 5010 2460 4755 1950 5010 2370 4800 1950
% 
\special{pn 4}%
\special{pa 4020 2600}%
\special{pa 3495 1550}%
\special{fp}%
\special{pa 4065 2600}%
\special{pa 3540 1550}%
\special{fp}%
\special{pa 4110 2600}%
\special{pa 3585 1550}%
\special{fp}%
\special{pa 4155 2600}%
\special{pa 3630 1550}%
\special{fp}%
\special{pa 4200 2600}%
\special{pa 3675 1550}%
\special{fp}%
\special{pa 4245 2600}%
\special{pa 3720 1550}%
\special{fp}%
\special{pa 4290 2600}%
\special{pa 3765 1550}%
\special{fp}%
\special{pa 4335 2600}%
\special{pa 3810 1550}%
\special{fp}%
\special{pa 4380 2600}%
\special{pa 3855 1550}%
\special{fp}%
\special{pa 4425 2600}%
\special{pa 3900 1550}%
\special{fp}%
\special{pa 4470 2600}%
\special{pa 3945 1550}%
\special{fp}%
\special{pa 4515 2600}%
\special{pa 3990 1550}%
\special{fp}%
\special{pa 4560 2600}%
\special{pa 4035 1550}%
\special{fp}%
\special{pa 4605 2600}%
\special{pa 4080 1550}%
\special{fp}%
\special{pa 4650 2600}%
\special{pa 4125 1550}%
\special{fp}%
\special{pa 4695 2600}%
\special{pa 4170 1550}%
\special{fp}%
\special{pa 4740 2600}%
\special{pa 4215 1550}%
\special{fp}%
\special{pa 4785 2600}%
\special{pa 4260 1550}%
\special{fp}%
\special{pa 4830 2600}%
\special{pa 4305 1550}%
\special{fp}%
\special{pa 4875 2600}%
\special{pa 4350 1550}%
\special{fp}%
\special{pa 4920 2600}%
\special{pa 4395 1550}%
\special{fp}%
\special{pa 4965 2600}%
\special{pa 4440 1550}%
\special{fp}%
\special{pa 5010 2600}%
\special{pa 4485 1550}%
\special{fp}%
\special{pa 5010 2510}%
\special{pa 4530 1550}%
\special{fp}%
\special{pa 5010 2420}%
\special{pa 4575 1550}%
\special{fp}%
\special{pa 5010 2330}%
\special{pa 4620 1550}%
\special{fp}%
\special{pa 5010 2240}%
\special{pa 4665 1550}%
\special{fp}%
\special{pa 5010 2150}%
\special{pa 4710 1550}%
\special{fp}%
\special{pa 5010 2060}%
\special{pa 4755 1550}%
\special{fp}%
\special{pa 5010 1970}%
\special{pa 4800 1550}%
\special{fp}%
% LINE 3 0 3 2
% 8 5010 2280 4845 1950 5010 2190 4890 1950 5010 2100 4935 1950 5010 2010 4980 1950
% 
\special{pn 4}%
\special{pa 5010 1880}%
\special{pa 4845 1550}%
\special{fp}%
\special{pa 5010 1790}%
\special{pa 4890 1550}%
\special{fp}%
\special{pa 5010 1700}%
\special{pa 4935 1550}%
\special{fp}%
\special{pa 5010 1610}%
\special{pa 4980 1550}%
\special{fp}%
\end{picture}%